Corrupted by Algorithms? How AI-generated and Human-written Advice Shape

(Dis)honesty

by

Margarita Leib*, Nils Köbis*, Rainer Michael Rilke, Marloes Hagens, & Bernd Irlenbusch

* shared first-authorship

This research has been approved by the Ethics Commission of the Faculty of Management, Economics, and Social Sciences of the University of Cologne under reference 200010BI.

**Acknowledgements:** We thank Clara Bersch, Yulia Litvinova, Ann-Kathrin Blanke, Toan Huynh, Anna Vogts and Matteo Tinè for research assistance, and Iyad Rahwan, Jean-Francois Bonnefon, Aljaz Ule, Anne-Marie Nussberger as well as the attendees of the Cognition, Values & Behaviour Research Group (Ludwig-Maximilians Universität München / LMU Munich), Moral AI lab meeting (Max Planck Institute for Human Development & Toulouse School of Economics), Applied Ethics & Morality Group (Prague University of Business and Economics), Centre for Decision Research (University of Leeds), Department of Economics and Management (University of Pisa), Decision Making and Economic Psychology Center (Ben-Gurion University), Behavioral and Management Science group (Technion), Colloquium of the Department of Social Psychology (Tilburg University) & Seminar at Department of Computer Science (Friedrich-Alexander University Erlangen-Nuremberg) for their helpful comments.

**Funding:** The research was funded by the Deutsche Forschungsgemeinschaft (DFG, German Research Foundation) under Germany's Excellence Strategy–EXC 2126/1–390838866' ECONtribute: Markets and Public Policy', the European Research Council (ERC-StG-637915), and the Chamber of Commerce and Industry (IHK) Koblenz.

**Abstract**

Artificial Intelligence (AI) increasingly becomes an indispensable advisor. New ethical concerns arise if AI persuades people to behave dishonestly. In an experiment, we study how AI advice (generated by a Natural-Language-Processing algorithm) affects (dis)honesty, compare it to equivalent human advice, and test whether transparency about advice source matters. We find that dishonesty-promoting advice increases dishonesty, whereas honesty-promoting advice does not increase honesty. This is the case for both AI- and human advice. Algorithmic transparency, a commonly proposed policy to mitigate AI risks, does not affect behaviour. The findings mark the first steps towards managing AI advice responsibly.

**Corrupted by Algorithms? How AI-generated and Human-written Advice Shape (Dis)honesty**

Artificial Intelligence (AI) shapes people's life on a daily basis (Rahwan et al., 2019). It sets prices in online markets (Calvano et al., 2020), predicts crucial outcomes such as healthcare costs (Obermeyer et al., 2019) and criminal sentences (Kleinberg et al., 2018), and makes recommendations ranging from entertainment content and purchasing decisions to romantic partners (Dellaert et al., 2020; Yeomans et al., 2019). Increasingly, AI has become an indispensable advisor, thereby affecting people's behaviour (Fast & Schroeder, 2020; Kim & Duhachek, 2020). As a case in point, Amazon's chief scientist, Rohit Prasad, envisions that Alexa's role for its over 100 million users "keeps growing from more of an assistant to an advisor" (Strong, 2020). Given AI's increasing role as an advisor, it is crucial to examine whether people are persuaded to follow or break ethical rules based on AI advice (Köbis et al., 2021).

Large companies like LinkedIn and Zillow are already implementing AI advisors, thereby potentially shaping their employees' ethical behaviour. In such companies, natural language processing (NLP) algorithms (e.g., provided by software such as Gong.io) analyse employees' recorded sales calls and advise them on how to increase their sales. Without supervision, such algorithms may detect that deceiving customers pays off and thus advise salespeople to do so. Indeed, NLP algorithms can already autonomously detect deception as a useful strategy in a negotiation task (Lewis et al., 2017). An ethical risk arises if people follow such corruptive AI advice. Here we examine (i) whether people meaningfully alter their (un)ethical behaviour following AI-generated advice and (ii) how such advice compares to human-written advice. Lastly, we test (iii) whether knowledge about the advice source (AI vs human) matters.

**Receiving advice on (un)ethical behaviour: Humans vs AI**

Generally, people are reluctant to take advice from others ("egocentric advice discounting", e.g., Yaniv & Kleinberger, 2000), especially when it is unsolicited (Bonaccio & Dalal, 2006). However, when facing an ethical dilemma, advice has several compelling benefits for the advised. Advice encouraging an ethical course of action may validate one's moral preferences. It thereby might reduce negative feelings such as regret for not taking the opportunity to maximise profits by lying. Advice encouraging an unethical course of action may free people to violate ethical rules for profit without spoiling their moral self-image (Cross et al., 2001). Indeed, taking advice can even provide a sense of shared responsibility with the advisor (Harvey & Fischer, 1997).

Compared to receiving human advice, how would people react to advice from an AI? Recent technological advances in the field of NLP reveal that AI text can already be indistinguishable from human text, suggesting AI advice is as convincing as human advice. For instance, GoogleDuplex, an AI-based call assistant, can book appointments while having full-fledged conversations without the recipient even realising that an AI is on the line. Further, AI can generate anything from poems (Köbis & Mossink, 2021) and Airbnb profiles (Jakesch et al., 2019) to news articles (Kreps et al., 2021) on par with humans. It thus stands to reason that when people are not informed about the sources of advice, they will not recognise the advice source correctly and be affected by AI and human advice similarly.

**Testing Algorithmic Transparency**

To make sure people know whom they interact with, governments, policymakers, and researchers univocally call for algorithmic transparency (Jobin et al., 2019) — the mandatory disclosure of AI presence (Diakopoulos, 2016). The recent Artificial Intelligence Act released by the EU demands AI systems such as chatbots and call assistants to disclose themselves as AI when interacting with humans (European Commission, 2021). Although it is a popular policy recommendation, empirical evidence for its effectiveness in shaping people's ethical behaviour is lacking.

How transparency about the advice source affects people's reaction to the advice is not trivial. Prior work informs three competing possibilities. The first possibility is that when informed about the source of advice, people follow human advice *more* than AI advice. This account rests on the literature on algorithm aversion (Dietvorst et al., 2015). People readily rely on AI in objective and technical domains (e.g., numeric estimation, data analysis, and giving directions, Castelo et al., 2019; Logg et al., 2019). However, they are reluctant to use AI for subjective decisions, especially with ethical implications (e.g., parole sentences, trolley-type dilemmas, Bigman & Gray, 2018; Castelo et al., 2019; Laakasuo et al., 2021). Further, people follow perceived social norms when making (un)ethical decisions (Bowles, 2016; Fehr, 2018; Gächter & Schulz, 2016; Gino et al., 2009; Köbis, Troost, et al., 2019). Compared to AI advice, human advice might be a stronger signal of social norms because social norms regulate and emerge from *human* (not AI) behaviour. Consequently, people should be more likely to follow human advice. Suppose people indeed prefer human input in ethically charged settings and perceive human advice as a stronger cue for social norms. In that case, we should expect that *human advice sways people's (un)ethical behaviour more than AI advice.*

The second possibility is that when informed about the source of advice, people follow advice from humans *less* than from AI. A closer look at the technical design of AI advice systems would support this account. NLP algorithms are trained on a large corpus of human-written texts (Radford et al., 2019). When people know that NLP algorithms draw on large compiled human input, they might perceive AI advice as a better representation of *most* people's beliefs and behaviours than the advice they receive from one human. If AI advice is indeed a stronger cue for social norms than a single piece of human-written advice, we should expect that *AI advice sways people's (un)ethical behaviour more than human advice.*

The third possibility is that when people receive information about the source of advice, they are affected *equally* by human and AI advice. Support for this account comes from the observation that people already seek advice from AI agents. For instance, more than 7 million people turn to Replika, the "AI companion who cares. Always here to listen and talk. Always on your side" (replika.ai) for virtual companionship, socialising, and also for advice (Murphy, 2019). Such AI advisors might also help justify questionable behaviour. When tempted to break ethical rules for profit, people do so as long as they can justify their actions (Barkan et al., 2015; Fischbacher & Föllmi-Heusi, 2013; Shalvi et al., 2015). Receiving advice that encourages rule-breaking can serve as a welcomed justification, possibly even when the advice stems from AI. Indeed, people deflect blame and share the responsibility for harmful outcomes not only with other people (Bartling & Fischbacher, 2011; Bazerman & Gino, 2012; Tenbrunsel & Messick, 2004) but also with AI systems (Hohenstein & Jung, 2020). If following AI and human advice is equally justifiable and leads to similar attribution of responsibility between the two, we should expect that *human and AI advice sway people's (un)ethical behaviour to the same extent.*

**The current study**

The current study tests how advice type (honesty- vs dishonesty-promoting), advice source (AI vs Human), and information about advice source (transparency vs opacity) shape humans' (un)ethical behaviour. Prior work has examined people's *stated preferences* about *hypothetical scenarios* describing AI advice (Bigman & Gray, 2018; Castelo et al., 2019; Kim & Duhachek, 2020; Logg et al., 2019). We supplement such work by adopting a machine behaviour approach (Rahwan et al., 2019) and examine people's *behavioural reactions* to *actual AI-generated output*.

To measure people's (un)ethical behaviour, we use the well-established incentivised die-rolling task (Fischbacher & Föllmi-Heusi, 2013). In it, participants roll a die in private and report the outcome, with higher outcomes corresponding to higher pay (see for similar approaches, Abeler et al., 2019; Fehrler et al., 2020; Sutter, 2008). To generate real AI advice, we employ the state-of-the-art algorithm GPT-J. We fine-tune the algorithm using minimal training to produce relevant advice texts. Using this "few-shot" learning approach mimics many real-world settings where language models (e.g., Alexa) are not optimised to maximise users' profits or provide the most convincing advice but simply produce relevant information based on prompts (Brown et al., 2020).

**Method**

We conducted a large-scale, pre-registered experiment. The design entailed two parts. In part 1, we collected human-written advice and generated AI advice; in part 2, we examined the effect of advice on (dis)honest behaviour (see Figure 1 for an overview). By using real AI text outputs, we avoid experimental deception and can gain insights into how people react to actual AI advice. We report all measures, manipulations, and data exclusions in the main text and Appendix. The IRB board of our department approved the experiment, and all materials, pre-registrations, and data are available on the Open Science Framework.

### Part 1 - Advice-giving task

**Human-written advice.** First, we conducted an *advice-giving task* in which we recruited advisors ($N$ = 367, $M_{age}$ = 30.28, $SD_{age}$ = 9.56, 45.50% females) via Prolific.co (pre-registration https://osf.io/nbke2). The task took 10-15 minutes, and participants earned a base pay of £1.27. Advisors learned that a separate group of participants (advisees) would engage in a die-rolling task (Fischbacher & Föllmi-Heusi, 2013), in which they roll a die privately and report the outcome (with higher outcomes corresponding to higher pay). Advisors were asked to write advice for these advisees.

Advisors were randomly assigned to either write honesty-promoting or dishonesty-promoting advice. We informed advisors in the *Honesty-promoting* advice treatment that if an advisee read their advice and subsequently reported the actual die-roll outcome (i.e., acted honestly), they would earn a bonus of £3. Advisors in the *Dishonesty-promoting* advice treatment learned that if an advisee read their advice and subsequently reported the highest possible outcome, 6, they would earn a bonus of £3[1].

Advisors had to follow pre-specified advice writing rules to ensure they produced coherent advice texts that could be used to train GPT-J. Specifically, their advice had to (i) entail at least 50 words, (ii) not use concrete numbers in numeric or written form[2], (iii) be in English and in their own words, (iv) be written in complete sentences, (v) be about the advisee's die-roll outcome reporting decision, and (vi) not inform the advisee that the advisor's payoff depended on their behaviour[3].

---

[1] If advisees follow the advice in the dishonesty-promoting treatments, they will lie in the majority of the cases (5 out of 6 cases). Only when the actual die-roll outcome is 6, following the advice does not entail lying.

To incentivise advisors to follow the advice writing rules, they stood to gain a bonus. Namely, out of all advice texts, we randomly selected one, and if that text followed the writing rules, the advisor earned a bonus of £10. Moreover, as incentivisation for writing convincing texts, 1 per cent of advice texts (4 out of 400) were implemented. If advisees acted according to the implemented advice, the respective advisor earned a bonus based on the treatment they were in (*Honesty-* vs *Dishonesty-promoting* advice)[4].

**AI-generated advice.** To generate AI advice (see Figure 1A), we employed GPT-J[5], an open-source NLP algorithm published by Eleuther AI (https://www.eleuther.ai/). GPT-J is trained on a curated and diverse data set of 825 GiB texts to predict the next word in a sequence of words and contains 6 billion parameters (Wang & Komatsuzaki, 2021). GPT-J can be fine-tuned with extra training to produce a specific type of text. We fine-tuned GPT-J with "few shot" learning by separately training it on the human-written honesty-promoting and dishonesty-promoting advice from the *advice-giving task.* We only used advice texts that adhered to the advice writing rules (as coded by a naive coder) for fine-tuning. More details on the calibration of GPT-J are reported in the Appendix.

---

[2] Advisors were not allowed to use concrete numbers to allow generating high-quality AI advice. GPT-J is trained to predict the next word in a sentence (see 'AI-generated advice' section). If advisors were allowed to concretely mention numbers, training GPT-J on the human written advice could have resulted in random numbers appearing out of context in the GPT-J output, reducing the quality of AI-generated advice.

[3] Advisors were not allowed to mention their incentive structure to the advisees so that we could keep the prosocial motivation for advisees who read AI and human advice constant (at zero).

[4] Paying advisors required knowing whether participants, after reading the advice, reported the observed die-roll honestly or not. To do so, we ran a modified version of the die-rolling task in which advisees received randomly selected advice, saw a die-roll on the computer screen and were asked to report it. We implemented this procedure for four randomly selected advice texts (1% of the advice) and four advisees. This non-private procedure provided certainty about whether an advisee reported honestly or not and enabled us to pay advisors accordingly. Doing so meant that our experimental setup was incentivised and did not entail experimental deception. In the main experiment, the die-roll outcomes were private (see 'Part 2 - Advice-taking task').

[5] As one can read in our pre-registration, we originally planned on deploying GPT-2 (see https://openai.com/blog/better-language-models/) to generated AI advice. However, we opted to use GPT-J instead because it is open source, which increases reproducibility and is more advanced as it is much larger and more potent than GPT-2.

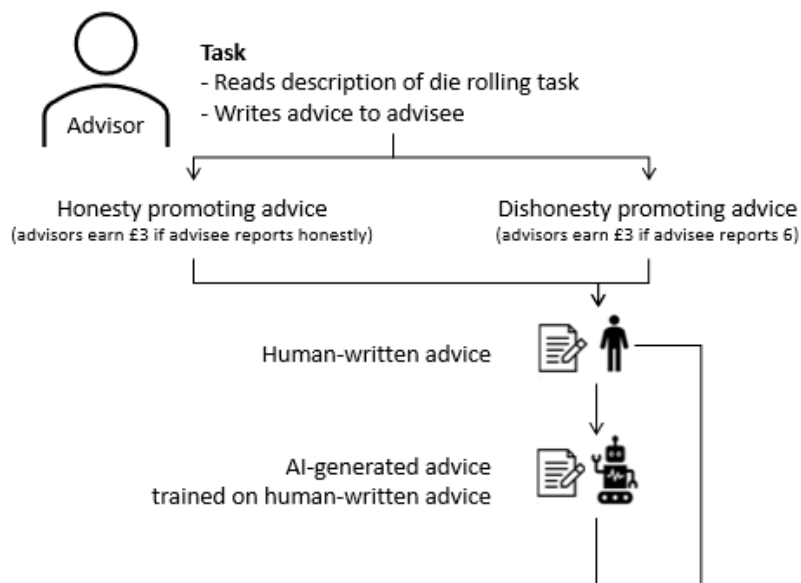

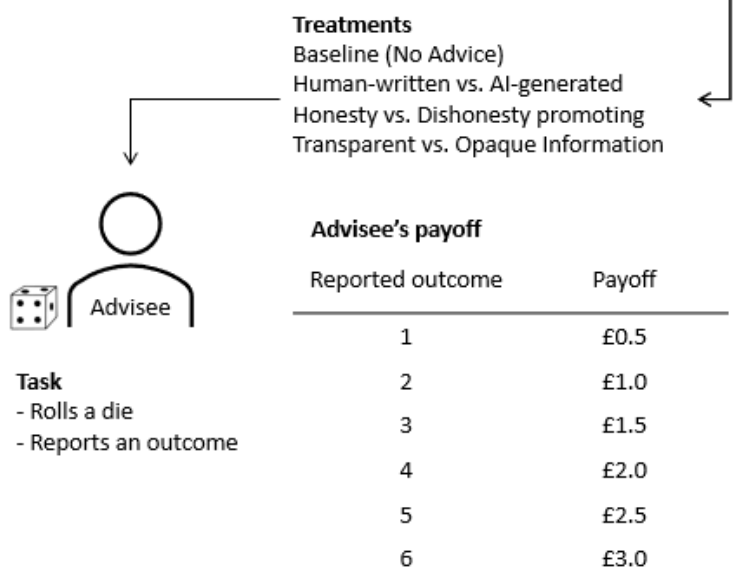

**Figure 1**. (A) Part 1 - advice-giving task (B) Part 2 - advice-taking tasks. (A) Participants were incentivised to write honesty- or dishonesty-promoting advice texts, which were then used to generate AI advice. (B) Another group of participants engaged in the die-rolling task. Advisees read advice, then reported a die-roll outcome. In total, we administered nine treatments: Participants read honesty or dishonesty-promoting advice that was human-written or AI-generated. Participants were either informed about the source of advice (Transparency) or not (Opacity). As a baseline, another group of participants did not read any advice.

  **Screening.** After collecting human advice and generating AI advice, we employed the same pre-specified screening procedure for both sources (see Figure 2). First, we excluded texts that exceeded 100 words. Next, to ensure advisees read coherent and relevant advice texts, we randomly selected 100 advice texts per cell. Two independent coders, who were naive to the experimental treatments, coded each piece of advice on the following criteria: (a) is the text coherent? (Y/N); (b) does the text contain clear advice? (Y/N); (c) which type of behaviour does the advice encourage? (honesty/dishonesty/unclear); (d) does the advice follow advice writing rules? (Y/N). Further, we used the objective Grammarly and Readability scores as computational proxies for the quality of the texts[6].

  Among the texts that passed the coding procedure[7] and received a Grammarly score equal or above 50, we randomly selected 20 advice texts per treatment (*AI-generated* vs *Human-written*, by *Honesty-* vs *Dishonesty-promoting*), yielding a final sample of 80 advice texts used in part 2 (see all advice texts in the Appendix). By applying the same screening procedure for human and AI advice, we ensure that the advice texts fulfil minimal quality criteria and are as comparable as possible.

---

[6] We obtained Grammarly and Readability scores from grammarly.com. Grammarly score compares texts to all other texts checked on the platform. A score of 80 indicates that a text scores better than 80% of all texts checked on grammarly.com in terms of grammatical correctness. Readability score employs the Flesch-Reading-ease test and represents how easy a text is to read. The score is calculated by the average sentence length and the average number of syllables per word, with higher scores indicating easier readability.

[7] Texts that passed the coding procedure (i) are coherent, (ii) contain clear advice, (iii) encourage honesty in the honesty-promoting treatment and dishonesty in the dishonesty-promoting treatment, and (iv) follow the advice writing rules. Moreover, the coding by both independent coders had to match each other in order for the text to pass.

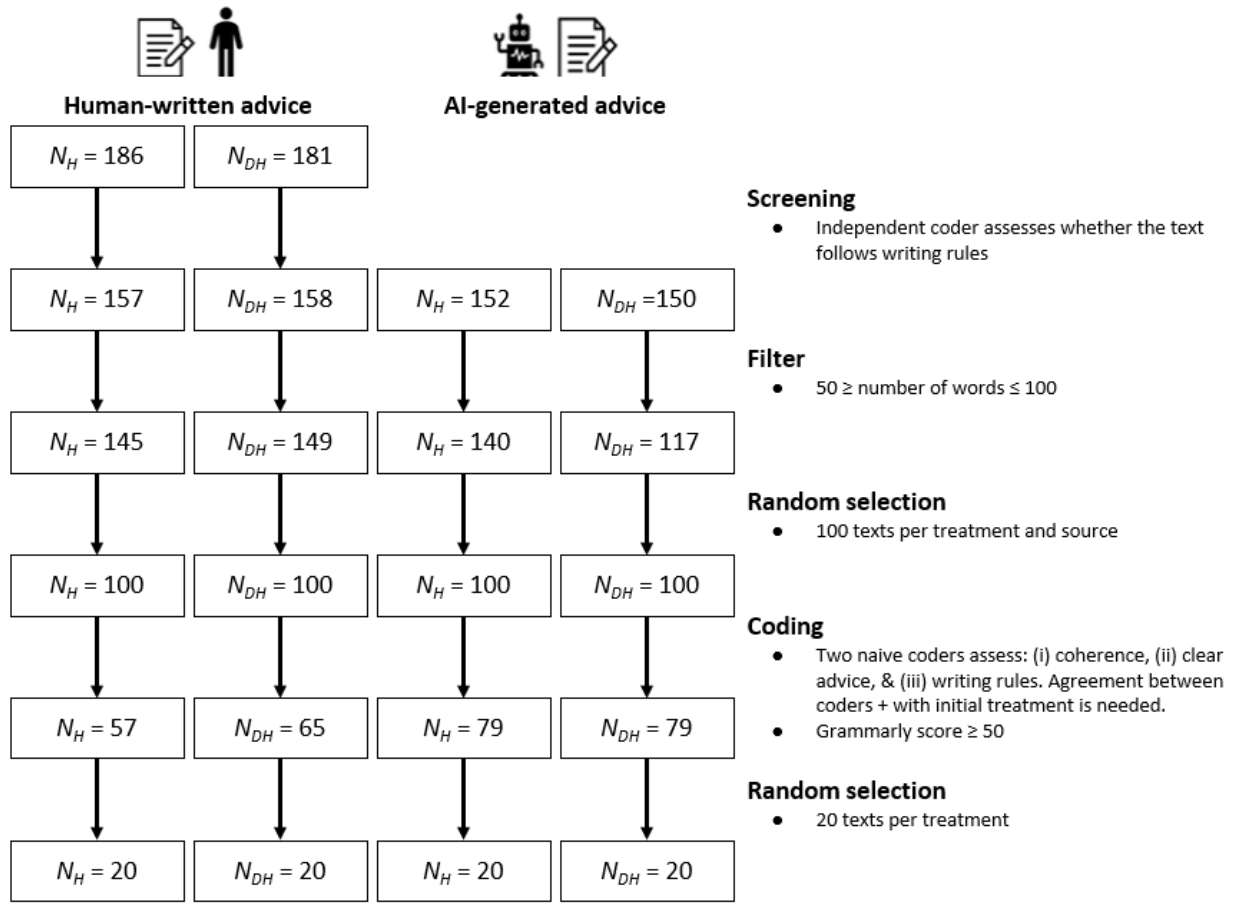

**Figure 2.** Overview of the selection procedure of advice texts. H = Honesty-promoting advice; DH = Dishonesty-promoting advice.

### Part 2 - Advice-taking task

The advice-taking task took about 8 minutes to complete, and participants earned a fixed pay of £1.20. We pre-registered (https://osf.io/nqvf3) to collect a sample size that would allow us to detect a small to medium effect size (200 participants per cell, 1,800 in total) via Prolific.co to take part in the *advice-taking task*. Overall, 1,817 ($M_{age}$ = 32.39; $SD_{age}$ = 11.68, 48.72% females) participants were included in the analyses. These participants completed the task and self-report items and passed the comprehension and attention checks (see below). Sensitivity analysis for a regression with 90% power and a significance level of .05 revealed our sample was sufficient to detect small effect sizes ($f^2$ = .006 and .010, see Appendix for details).

Participants read the instructions, received advice, and finally engaged in the die-rolling task. Specifically, participants were asked to roll a die privately and report the outcome. Participants' pay corresponded to their report: for reporting a '1' they earned £0.5; for a '2' = £1, '3' = £1.5, '4' = £4, '5' = £2.5, '6' = £3. After reading the instructions and before engaging in the die-rolling task, all participants learned that 10 per cent of participants would be randomly selected and paid for the die-rolling task.

Assessing dishonesty by employing the die-rolling task is a common practice in economics and psychology (see meta-analyses, Abeler et al., 2019; Gerlach et al., 2019; Köbis, Verschuere, et al., 2019; Leib et al., 2021). Further, the task has good external validity, as lying in the die-rolling task correlates with unethical behaviour outside the lab, such as free-riding public transportation (Dai et al., 2018) and being absent from work without reason (Hanna & Wang, 2017). Importantly, the die-rolling task pits two competing motivations against each other: to be a moral, honest person vs maximising financial profits, thus rendering advice particularly valuable.

After reading the instructions, participants had to answer three comprehension questions correctly and pass an attention check. If participants answered a comprehension question incorrectly, they could try again. If they answered the same question incorrectly twice, or if they answered the attention check incorrectly, they were disqualified from participating in the study and were not paid.

**Treatments.** Before reporting the die-roll outcome, participants were randomly assigned to one of nine different treatments. Participants either (i) read *Honesty-promoting* or *Dishonesty-promoting* advice, (ii) that was either *Human-written* or *AI-generated*. Further, (iii) participants were either informed about the advice source (*Transparency* treatments) or not (*Opacity* treatments). Additionally, in a *Baseline* treatment, participants did not receive any advice. Thus, the overall design was a 2 (Advice source: Human vs AI) by 2 (Advice type: Honesty-promoting vs Dishonesty-promoting) by 2 (Information about the advice source: Transparency vs Opacity) + 1 (Baseline, no advice) between-subject design.

Participants who read AI-generated advice and were informed about it (in the *Transparency* treatments) read:

> "To help you with your decision, an artificially intelligent (AI) algorithm generated an advice.
>
> <u>Information about AI-advice:</u>
>
> The algorithm is trained on large datasets of English text (> 820 GiB, which is more than 500 million pages of text) and produces text by predicting the next word in a sentence (similar to predictive text on smartphones). On top of that, it is also possible to train the algorithm on additional specific pieces of text.
>
> To generate multiple AI advice texts, the algorithm was trained on advice texts collected from other participants in the study. These participants did not take part in the die rolling task and were only instructed to write advice regarding the decision in

> *the die rolling task. The advice you will read is one advice text that was generated by*
>
> *the algorithm."*

Participants who read human-written advice and were informed about it (in the *Transparency* treatments) read:

> *"To help you with your decision, another participant wrote an advice.*
>
> <u>*Information about advice:*</u>
>
> *To collect multiple advice texts, another group of participants was asked to write*
>
> *advice regarding the decision in the die rolling task. These participants did not take*
>
> *part in the die rolling task and were only instructed to write advice regarding the*
>
> *decision in the die rolling task. The advice you will read is advice written by one*
>
> *participant. "*

Participants who were in the *Opacity* treatments and thus not informed about the advice source read:

> *"To help you with your decision, you will read an advice.*
>
> *This advice has been written either by another participant or by an artificially intelligent (AI)*
>
> *algorithm. There is a 50% chance the advice is written by a participant and a 50% chance it is*
>
> *written by an algorithm."*

In the *Opacity* treatments, this text was followed by the same two descriptions of how advice text from each source was collected or generated in the *Transparency* treatments. In the *Opacity* treatment, this information about AI advice generation and human advice collection appeared in random order.[8]

---

[8] To control for participants' beliefs about the potential advice sources, we opted to inform them that there is a 50-50 chance that a human or AI wrote the advice. We believed that not providing any information about the advice source would reasonably lead participants to assume the advice source is another human, as AI might not be a salient source of advice for participants.

**A static Turing test.** After completing the die-rolling task, participants in the *Opacity* treatment engaged in an incentivised version of a static Turing Test (Köbis & Mossink, 2021). In contrast to the classical Turing Test (Turing, 1950), participants did not interact back and forth with the source of advice. Instead, they read the advice text and indicated whether they thought a human or an AI had written it. Participants learned that 20 of them would be randomly selected, and if their guess in the static Turing test was correct, they would earn an additional £1.

**Potential mechanisms.** Finally, to explore possible mechanisms, participants completed a post-experimental survey. Participants indicated on a scale from 0 to 100 their perceived (i) appropriateness (injunctive social norm), (ii) prevalence (descriptive social norm), and (iii) justifiability of reporting a higher die-roll than the one observed. Additionally, all participants, except those who did not receive any advice, rated how they attribute responsibility between themselves and the advisor for the reported outcome in the die-rolling task. The answer scale ranged from 0 (= I am fully responsible) over 50 (= The advisor and I share responsibility equally) to 100 (= The advisor is fully responsible). Participants further indicated (on a scale from 0 to 100) to what extent they feel guilty after completing the task (see Appendix for results regarding guilt and wording of all items). Finally, all participants indicated their age and gender.

## Results

In all nine treatments, participants lied as the average die-roll outcomes significantly exceeded the expected average if participants were honest (EV = 3.5), one-sample *t*-test, $ts > 3.43$, $ps < .001$.

### *Is people's (un)ethical behaviour influenced by AI-generated advice?*

Yes, when it comes to dishonesty-promoting advice; no, when it comes to honesty-promoting advice. We first focus on the *Opacity* treatments, where participants are not informed about the advice source. Here, linear regression analyses reveal that the average die-roll reports following *AI-generated Dishonesty-promoting* advice (*M* = 4.59, *SD* = 1.37) significantly exceed reports in the *Baseline*, no advice treatment (*M* = 3.98, *SD* = 1.55, *b* = .609; *p* < .001; 95% CI = [.324, .894]). However, die-roll reports following *AI-generated Honesty-promoting* advice (*M* = 4.00, *SD* = 1.62) do not significantly differ from reports in the *Baseline* treatment (*b* = .019; *p* = .898; 95% CI = [-.275, .314]), see Figure 3 and Table 1 (model 1). Further, die-roll reports in the *AI-generated Dishonesty-promoting* treatment significantly exceed those in the *AI-generated Honesty-promoting* advice treatment (*b* = -.590, *p* < .001; 95% CI = [-.881, -.299]). Thus, while dishonesty-promoting AI advice successfully corrupts people, honesty-promoting AI advice fails to sway people toward honesty.

### How does AI-generated advice square compared to human-written advice?

AI-generated advice affects behaviour similarly to human-written advice, for both honesty-promoting and dishonesty-promoting advice. Focusing on the *Opacity* treatments, the two-way interaction (advice source by advice type) is not significant (*b* = .069, *p* = .744; 95% CI = [-.349, .489]), see Figure 3 and Table 1 (model 2). Specifically, the average die-roll reports do not differ between the *AI-generated* (*M* = 4.00, *SD* = 1.62) and *Human-written* advice when advice was *Honesty-promoting* (*M* = 3.92, *SD* = 1.51, *b* = -.076, *p* = .631; 95% CI = [-.387, .235]). Similarly, average die-roll reports do not differ between the *AI-generated* (*M* = 4.59, *SD* = 1.37) and *Human-written* advice when advice was *Dishonesty-promoting* (*M* = 4.58, *SD* = 1.53, *b* = -.006, *p* = .964; 95% CI = [-.289, .276]).

In addition, the results of the static version of the Turing Test indicate that individuals cannot distinguish AI-generated advice from human-written advice. Specifically, in the *Opacity* treatments, 49.93 per cent (401 out of 803) of participants guessed the source of advice correctly, which does not differ from chance levels (50%, binomial test: $p$ = .999; 95% CI = [.464, .534]).

### *Does transparency about the advice source matter?*

No, informing participants about the algorithmic or human source of advice does not change their behaviour. Linear regression analyses reveal that the three-way interaction (advice type by source by information) is not significant ($b$ = .100, $p$ = .735; 95% CI = [-.482, .683]), Figure 3 and Table 1 (model 3). Both among the *Opacity* and *Transparency* treatments, the two-way interactions (advice source by advice type) are not significant (*Transparency*: $b$ = .170, $p$ = .409, 95% CI = [-.234, .575]; *Opacity*: $b$ = .069, $p$ = .744, 95% CI = [-.349, .489]).

Overall, the popular policy recommendation of algorithmic transparency does not alleviate the corrupting effect of AI advice. Namely, die-roll reports following *AI-generated Dishonesty-promoting* advice under the *Opacity* treatment ($M$ = 4.59, $SD$ = 1.37) are on par with reports following the same advice in the *Transparency* treatment ($M$ = 4.61, $SD$ = 1.40, $b$ = .020, $p$ = .878; 95% CI = [-.244, .286]). Specifically, when participants are *not informed* about the advice source, they boost their reports by 15.2% following *AI-generated Dishonesty-promoting* advice, compared to the *Baseline* [(4.59-3.98)/3.98 = .152], which is equivalent to the 15.2% increase when they *are informed* about the source of the advice [(4.61-3.98)/3.98 = .158]. Bayesian analyses corroborate these conclusions (see Appendix).

Overall, results align with the idea that people increasingly follow AI advice (e.g., Replika) and use AI-generated advice to justify breaking ethical rules for profit.

**Robustness of the obtained results.** In our experimental design, advisors in the *Dishonesty-promoting* treatment received £3 only if advisees reported the highest value, '6'. Such an incentive scheme is comparable with the *Honesty-promoting* treatment in which advisors earned £3 only if advisees reported honestly. In both cases, advisors earn money for 1 out of 6 potential advisee's reports (i.e., when the advisee reports' 6' or honestly, depending on the treatment) and do not earn money in the remaining 5 of the advisee's reports. However, advisors' incentive scheme in the *Dishonesty-promoting* treatments may have resulted in advice texts that predominantly focused on convincing participants to report the outcome 6. To assess the robustness of our results, we (i) conducted additional analyses and (ii) ran additional treatments.

*Proportion of sixes.* First, as an additional analysis, we examined whether the proportion of 6's, as an alternative outcome variable, led to the same conclusions. We found very similar results (see Figure 3, the white dots represent the proportion of 6's across all treatments). Specifically, focusing on the *Opacity* treatments, linear regression analyses reveal that the proportion of sixes following *AI-generated Dishonesty-promoting* advice (32.44%) significantly exceeds the proportion of sixes in the *Baseline*, no advice treatment (20.65%; $b$ = .612; $p$ = .005, 95% CI = [.182, 1.051]). However, the proportion of sixes following *AI-generated Honesty-promoting* advice (21.93%) does not significantly differ from the *Baseline* ($b$ = .076; $p$ = .751, 95% CI = [-.398, .551]). Further, the proportion of sixes in the *AI-generated Dishonesty-promoting* treatment significantly exceeds that in *AI-generated Honesty-promoting* advice treatment ($b$ = -.535, $p$ = .016, 95% CI = [-.979, -.101]).

Further, focusing on the *Opacity* treatments, the two-way interaction (advice source by advice type) is not significant ($b$ = .444, $p$ = .171, 95% CI = [-.191, 1.082]). The proportion of sixes does not differ between the *AI-generated* (21.93%) and *Human-written* treatments when the advice is *Honesty-promoting* (19.28%, $b$ = -.162, $p$ = .516, 95% CI = [-.654, .327]). Similarly, the proportion of sixes does not differ between the *AI-generated* (32.44%) and *Human-written* treatments when the advice is *Dishonesty-promoting* (38.91%, $b$ = .282, $p$ = .173, 95% CI = [-.123, 690]). Lastly, the three-way interaction (advice type by source by information) is also not significant ($b$ = -.523, $p$ = .257, 95% CI = [-1.430, .382]). Both among the *Opacity* and *Transparency* treatments, the two-way interactions (advice source by advice type) are not significant (*Transparency*: $b$ = -.078, $p$ = .810, 95% CI = [-.724, .566]; *Opacity*: $b$ = .444, $p$ = .171, 95% CI = [-.191, 1.082]).

**Additional (Aligned) treatments**. To assess the robustness of our results to the advisor's incentive scheme, we ran four additional treatments (advice source: *Human-written* vs *AI-generated* by information: *Transparency* vs *Opacity*). In these *Aligned* treatments, advisees read advice written by advisors whose incentives were aligned with those of the advisees. For these advisors ($n$ = 207), if the advisee reported '1', both the advisor and advisee earned £0.5 each; if the advisee reported '2', both the advisor and advisee earned £1 each and so on. We again fine-tuned GPT-J on such human-written advice texts. These treatments led to comparable results to the *Dishonesty-promoting* treatment. In particular, the average die-roll outcomes in all four *Aligned* treatments were significantly higher than in the *Baseline* treatment ($p$ = .066 for the *AI-generated, Opacity treatment*, and $p$s < .001 for the remaining three treatments, see Appendix for more details about these treatments and elaborated results). This consistency in results suggests that our results are robust to such variation in the advisors' incentive scheme.

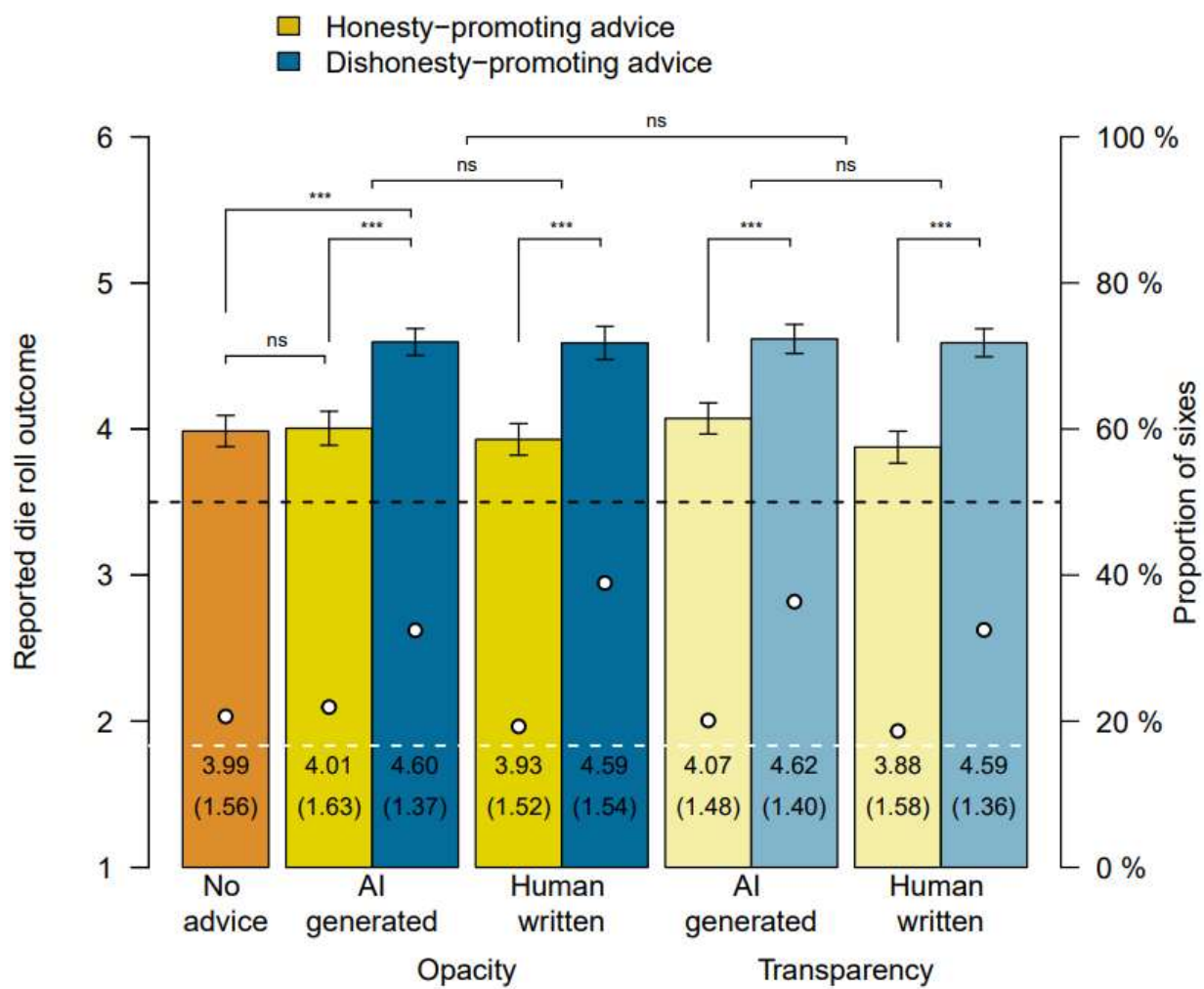

**Figure 3.** Mean reported die-roll outcomes (in bars) and proportion of reported 6s (in white dots) across advice type (honesty vs dishonesty-promoting), source (AI vs human), and information treatments (opacity vs transparency). The dashed black line represents the expected mean if participants were honest (EV = 3.5), and the dashed white line represents the expected proportion of 6s if participants were honest (16.67%). Mean (*SD*) of die-roll reports are at the bottom of each bar; ***$p < .001$; ns: $p > .05$.

**Potential mechanisms.** In line with the logic brought forth in the introduction, in this section, we examine whether participants' perception of (i) appropriateness (injunctive social norm), (ii) prevalence (descriptive social norm), and (iii) justifiability of reporting a higher die-roll than the one observed, as well as their (iv) attribution of responsibility between themselves and the advisor varies as a function of the advice source (AI vs human) and type (honesty vs dishonesty-promoting). Participants could not tell apart AI from human advice (indicated by the results of the static Turing test). Therefore, we focus only on treatments in which participants are informed about the advice source (*Transparency* treatments) to tap into the process of how known advice source and advice type shaped their perceptions. See the Appendix for the results of the *Opacity* treatment.

*Injunctive norms.* A linear regression predicting injunctive norms from the advice type (honesty vs dishonesty-promoting advice) revealed that participants evaluated reporting a higher die-roll outcome as more appropriate when reading a *Dishonesty-promoting* ($M$ = 33.93, $SD$ = 31.43) than *Honesty-promoting* advice ($M$ = 25.99, $SD$ = 29.68, $b$ = 7.94, $p$ < .001, 95% CI = [3.702, 12.182]). This finding indicates that the advice type shapes perceived injunctive norms. Notably, a linear regression predicting injunctive norms from advice type and source (AI vs human) revealed a non-significant advice source by type interaction, $b$ = -4.81, $p$ = .264, 95% CI = [-13.292, 3.658]. These results suggest that AI and human advice affected injunctive norms perceptions similarly (see Figure 4a). This result is consistent with the behavioural finding of participants' die-roll reports being affected by the type of advice but not by its source.

*Descriptive norms.* A linear regression predicting descriptive norms from the advice type revealed that participants evaluated reporting a higher die-roll outcome as more common when reading a *Dishonesty-promoting* (*M* = 76.02, *SD* = 22.75) than *Honesty-promoting* advice (*M* = 66.74, *SD* = 24.04, *b* = 9.27, *p* < .001, 95% CI = [6.031, 12.525]). This finding indicates that the advice type also shapes perceived descriptive norms. Importantly, a linear regression predicting descriptive norms from advice type and source revealed a non-significant advice source by type interaction, *b* = .25, *p* = .938, 95% CI = [-6.230, 6.745], indicating that AI and human advice affected descriptive norms perceptions similarly (see Figure 4b). This result is consistent with the behavioural finding, showing that advice type affected die-roll reports, but advice source did not.

     *Justifiability.* A linear regression predicting justifiability from the advice type revealed that participants evaluated reporting a higher die-roll outcome as more justifiable when reading a *Dishonesty-promoting* (*M* = 40.96, *SD* = 31.10) than *Honesty-promoting* advice (*M* = 28.45, *SD* = 28.25, *b* = 12.50, *p* < .001, 95% CI = [8.387, 16.629]). This finding suggests that the advice type shapes perceptions of how justifiable lying in the die-rolling task is. A linear regression predicting justifiability from advice type and advice source revealed a non-significant advice source by type interaction (*b* = -1.04, *p* = .803, 95% CI = [-9.280, 7.194]), indicating that AI and human advice affected justifiability perceptions similarly (see Figure 4c). This result is consistent with the behavioural finding, showing that the type of advice affected participants' die-roll reports, but the source of advice did not.

     *Shared responsibility.* The shared responsibility scale ranged from 0 (= I am fully responsible) to 100 (= The advisor is fully responsible), with 50 indicating equally shared responsibility between the participant and the advisor. On average, participants indicated they are more responsible for the outcome they report than the advisor (*M* = 27.59, *SD* =

36.60, one-sample *t*-test compared to the value 50, *t* = -17.32, *p* < .001). Further, a linear regression predicting shared responsibility from the advice source (AI vs human) revealed that participants attributed responsibility similarly when the advice source was an AI (*M* = 28.27, *SD* = 36.53) and human (*M* = 26.94, *SD* = 36.70, *b* = -1.326, *p* = .608, 95% CI = [-6.407, 3.754]). A linear regression predicting shared responsibility from advice type and advice source revealed a non-significant source-by-type interaction (*b* = -5.91, *p* = .253, 95% CI = [-16.083, 4.248], see Figure 4d). The fact that participants attribute responsibility between themselves and the advisor to the same extent regardless of whether the advisor is a human or an AI is consistent with the logic fleshed out in the introduction, in which people will follow human and AI advice similarly if they share responsibility with both advice sources to similar levels.

In sum, the results from the self-report items align with the third possibility outlined in the introduction. Namely, we find that participants' perceptions of injunctive and descriptive social norms and their perceived justifiability do not differ between human and AI advisors. Participants also attribute responsibility similarly between themselves and their advisor, regardless of whether the advisor is a human or an AI. This pattern of results mirrors the behavioural effects of AI and human advice affecting people's (dis)honesty similarly.

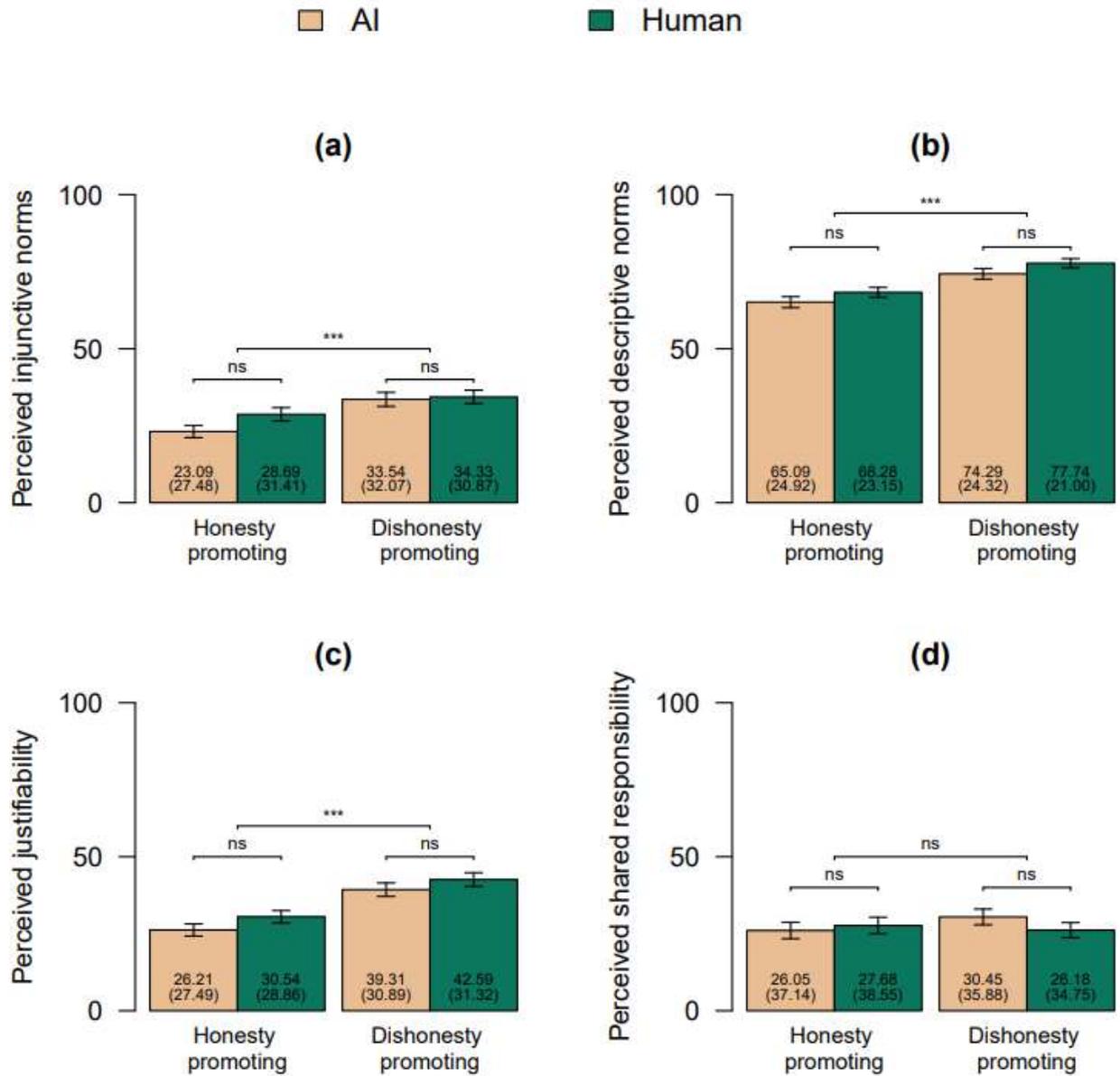

**Figure 4.** Mean reports of perceived (a) injunctive norms, (b) descriptive norms, (c) justifiability, and (d) shared responsibility across advice type (honesty vs dishonesty promoting) and source (AI [yellow] vs human [green]) in the transparency treatments. The means (*SD*) of reports are at the bottom of each bar; ***$p < .001$; ns: $p > .05$.

| | (1) | (2) | (3) | (4) | (5) | (6) | (7) |
|---|---|---|---|---|---|---|---|
| | | | | Dependent variable: Reported die-roll outcome | | | |
| No advice | .019 (.150) | | | | | | |
| Dishonesty-promoting advice | .609*** (.145) | .590*** (.147) | .590*** (.145) | .396** (.143) | .439** (.143) | .436** (.145) | .369* (.152) |
| Human-written advice | | -.076 (.152) | -.076 (.149) | -.166 (.146) | -.012 (.147) | -.024 (.156) | .086 (.170) |
| Transparency treatment | | | .067 (.150) | .071 (.146) | -.112 (.147) | .114 (.147) | |
| Dishonesty-promoting advice X Human advice | | .069 (.213) | .069 (.210) | .104 (.205) | .033 (.205) | -.045 (.209) | .041 (.218) |
| Dishonesty-promoting advice X Transparency treatment | | | -.046 (.208) | -.030 (.203) | -.088 (.204) | -.067 (.203) | |
| Human advice X Transparency treatment | | | -.120 (.210) | -.093 (.205) | -.145 (.206) | -.162 (.205) | |
| Dishonesty-promoting advice X Human advice X Transparency treatment | | | .100 (.297) | .082 (.289) | .169 (.290) | .160 (.290) | |
| Injunctive norms | | | | .001 (.001) | .001 (.001) | .001 (.001) | .003 (.002) |
| Descriptive norms | | | | .006** (.001) | .005** (.001) | .005** (.001) | .005* (.002) |
| Justifiability | | | | .007*** (.001) | .007*** (.001) | .007*** (.001) | .005* (.002) |
| Shared responsibility | | | | -.001 (.001) | .001 (.001) | .001 (.001) | .001 (.001) |
| Gender (male) | | | | | .188** (.072) | .191** (.072) | .203+ (.105) |

| | (1) | (2) | (3) | (4) | (5) | (6) | (7) |
|---|---|---|---|---|---|---|---|
| Age | | | | | -.008* (.003) | -.008* (.003) | -.005* (.004) |
| Grammarly score | | | | | | .007* (.003) | .013* (.005) |
| Readability score | | | | | | -.006 (.003) | .004 (.005) |
| Correctly guessed the source (1) or not (0) | | | | | | | .163 (.105) |
| Intercept | 3.98*** | 4.00*** | 4.00*** | 3.35*** | 3.57*** | 3.36*** | 1.83* |
| $R^2$ | .034 | .041 | .044 | .095 | .100 | .104 | .105 |
| $N$ | 634 | 1016 | 1604 | 1604 | 1593 | 1593 | 798 |
| Data used for analysis | Opacity, AI advice & Baseline, no advice | Opacity | All treatments without Baseline no advice | All treatments without Baseline no advice | All treatments without Baseline no advice | All treatments without Baseline no advice | Opacity Without Baseline no advice |

**Table 1.** Regression analyses on the average die-roll reports, including control variables and interactions. Models 5-7 contain a smaller $N$, as some participants did not report their gender as male/female. $+p < .10$, $*p < .05$, $**p < .01$, $***p < .001$.

## Discussion

As intelligent machines take an ever-growing role as advisors (Rahwan et al., 2019), and adherence to ethical rules crucially impacts societal welfare (Gächter & Schulz, 2016), studying how AI advice influences people's (un)ethical behaviour bears immense relevance (Köbis et al., 2021). We find that people follow AI-generated advice that promotes dishonesty, yet not AI-generated advice that promotes honesty. In fact, people's behavioural reactions to AI advice are indistinguishable from reactions to human advice. Substantiating that current-day NLP models can produce human-like texts, participants in our experiment could not tell apart human-written from AI-generated advice texts.

We further tested the commonly proposed policy of algorithmic transparency (Jobin et al., 2019) as a tool to mitigate AI-associated risks. Specifically, we examine whether knowing the source of the advice impacts people's reactions to it. The policy rests on the assumption that people adjust their behaviour when they learn that they interact with AI systems and not humans. Our experiment tested this assumption and revealed that algorithmic transparency is insufficient to curb AI advice's corruptive influence. Knowing that a piece of advice stems from an AI does not make people less (or more) likely to follow it compared to human-written advice.

Tapping into the mechanisms underlying these behavioural results, participants perceived lying as equally acceptable, common and justifiable when humans or AI promoted such dishonest behaviour. They further attribute responsibility similarly to AI and human advisors. These perceptions are consistent with previous work showing that in ethical dilemmas, people rely on justifications (Shalvi et al., 2015) and social norms (Abbink et al., 2018) and, by now, blame not only humans but also AI systems for adverse outcomes (Hohenstein & Jung, 2020). Advancing the justified ethicality theory, we, therefore, show that (i) dishonesty-promoting advice serves as a justification and social norms signal and (ii) that such advice does not even have to come from a human but can also be crafted by an AI.

In our setting, we collected human-written advice, created AI-generated advice, and then implemented a screening procedure for both human and AI advice to ensure that all advice texts are coherent, clear, and of decent quality. Such screening procedure allowed us to examine how *comparable* AI and human advice shape people's ethical behaviour and whether information about the advice source matters. Harmonising the quality of the texts allowed us to eliminate the alternative explanation that variations in text quality drive the obtained results. At the same time, the screening process introduced a human component to AI advice. Put differently, humans – in our case, naive coders – were "in the loop" of AI advice text generation. Note that 79 per cent of AI advice passed the quality screening criteria, while for human text, this passing rate was 57 and 65 per cent (honesty-promoting and dishonesty-promoting advice, respectively; Figure 2). These high screening passing rates for AI-generated texts demonstrate that current NLP algorithms can produce good-quality advice text without much prior training and optimisation.

Interesting extensions of our work could test the lower and upper limits of the effects of AI advice on ethical behaviour. To test the lower limit of the effect, future work can relax human control over the generation of AI advice. For instance, not implementing a screening procedure, thus removing humans "from the loop" when generating AI advice, will allow examining how unconstrained texts affect humans' behaviour (see for similar methodology, Köbis & Mossink, 2021). To test the upper limit of the effect, future work can examine AI's learning abilities to write convincing advice. One could use reinforcement learning to train an algorithm over multiple rounds of advice-giving, providing feedback after every written piece of advice. To obtain a symmetric comparison to humans' learning abilities, human advisors could similarly receive feedback after each piece of advice they write (see for a similar approach, Koster et al. 2022).

Previous work has documented a general *stated* aversion towards AI advice, with only 8% saying they would trust mortgage advice from AI (similar to the 9% who trust investment "advice" from a horoscope, HSBC, 2018). However, our behavioural results paint a different picture. In line with the growing practice of turning to AI agents such as Replika or Alexa for companionship and advice (Fast & Schroeder, 2020; Murphy, 2019), we find that people willingly adopt advice from AI when it aligns with their preferences. Our results indicate a discrepancy between individuals' *stated* preferences and *actual* behaviour, highlighting the importance of complimenting work on stated preferences with work adopting a machine behaviour approach – the study of human behaviour in interaction with real algorithmic outputs (Rahwan et al., 2019).

The process through which employing AI advice can result in humans' ethical rule violations consists of two main steps. The first step is algorithms being programmed on a certain objective function (e.g., maximising profits) that results in a (maybe unintended) corruptive advice. Indeed, NLP algorithms already detect and use deception as a useful strategy in a negotiation task (Lewis et al., 2017). The second step is people being affected by such corruptive AI advice. Practically, AI advice poses an ethical risk only if humans actually follow it. The current work focuses on this second step, showing that corruptive AI advice indeed poses an ethical risk, because people follow it to the same extent as human corruptive advice. We hope the current work can be of use to AI programmers (e.g., by preventing AI from bluntly advising unethical courses of action). More importantly, we call for more work from social scientists testing successful interventions that prevent people from following (AI) advice when it encourages unethical behaviour thereby mitigating its corruptive force.

## Conclusion

People increasingly use and interact with AI, which can provide them with unethical advice. Anecdotally, we asked a newly created Replika for advice regarding the ethical dilemma presented in the current experiment. Replika first provided rather vague advice ("If you worship money and things (...) then you will never have enough"), but when asked whether it prefers money over honesty, it replied: "money." We find that when faced with the trade-off between honesty and money, people will use AI advice as a justification to lie for profit. As algorithmic transparency is insufficient to curb the corruptive force of AI, we hope this work will highlight, for policymakers and researchers alike, the importance of dedicating resources to examining successful interventions that will keep humans honest in the face of AI advice.

Corrupted by Algorithms? How AI-generated and Human-written Advice Shape (Dis)honesty

**Appendix**

**Contents**

## GPT-J model used for AI-generated advice

To generate AI advice, we employed GPT-J, a natural language processing (NLP) algorithm that is trained to predict the next word given all the previous words within a text (Wang & Komatsuzaki, 2021). GPT-J contains 6 billion parameters and is trained on a diverse data set called "The Pile", a diverse, high-quality, and curated 825 GiB dataset (open source) that is used for language modelling purposes (Gao et al., 2020). GPT-J can be used for multiple language tasks, but the model is best at what it was pre-trained for: generating text from a prompt. GPT-J can further be fine-tuned with an extra training set. We fine-tuned GPT-J separately on the human-written (a) honesty-promoting and (b) dishonesty-promoting advice texts from the *advice-giving task.* The model, including training data, code and a fine-tuning guide (Wang, 2021), is available online to facilitate the reproduction of the algorithmic outputs. We used an API (Forefront) for fine-tuning the model and generating the texts. The human-written texts that were used to train the GPT-J model are available upon request.

Due to the often inexplicable "black box" nature of GPT-J, deciding how much fine-tuning is needed for our application of the GPT-J model is not trivial. We aimed to strike a balance between avoiding overfitting the GPT-J model (and thus having GPT-J generate output that is almost identical to the input) and underfitting it (and thus having GPT-J generate "gibberish"). To evaluate the model's performance, we used checkpoints. We had five checkpoints and picked checkpoint 4 for the final usage. Also, we used conditional sample generation because this generated higher-quality sequences. That is, we provided GPT-J with a specific text as a starting point (a prompt) and let GPT-J generate the entire advice text *after* the prompt. More specifically, we first fine-tuned the GPT-J algorithm with a prompt text before each of the selected human-written pieces of advice. Each human-written text started with the following instruction text: "Instruction: Write advice for the die-rolling game\n\nAdvice:". The actual advice of the participant ensued this prompt. Next, to generate advice texts, we also prompted the algorithm with the same prompt but left the actual advice blank and let the model complete it.

Further, we set the GPT-J model's parameters to be: (1) temperature = 0.9: float value controlling the randomness in the Boltzmann distribution; as the temperature value approaches zero, the model will become deterministic and repetitive (less random), whereas higher temperature values (e.g., 1) results in more random completions. Temperature values can range between 0 and 1. (2) length = 150: Texts contain, on average, 150 tokens (some generated sequences were a bit shorter than 150 tokens). The length refers to the length of the generated text, in tokens, based on the prompt. One token is approximately four characters. Using 150 tokens as a setting ensured us to get long enough output sequences. (3) top-$p$ = 1: float value controlling diversity and implements nucleus sampling. Top-$p$ values range between 0 and 1. When the top-$p$ is set to a float <1, *only the most probable* tokens that add to the top-$p$ value (probability) are kept for text generation. A lower value of top-$p$ means that the tokens returned will be more likely (or more 'safe'), whereas a higher value of top-$p$ means that the tokens returned will be more creative. (4) repetition penalty = 1.1: float values that represent a penalisation for repeated words, with higher values indicating the model is more penalised for repeating words. The default of 1 means that there is no penalisation. Depending on the task at hand, the value typically ranges between 1.1 and 1.3. Potentially, one can set it lower than 1 to increase repetitiveness. (5) top-$k$ = 30: integer value controlling diversity and restricts how many words are considered at each step (1 = only one word is considered at each step, resulting in deterministic completions; 50 = 50 words are considered at each step). Top-$k$ values range between 1 and 50. For all the other parameter values, we used the default settings. To find more information on transformers (such as GPT-J), including an overview of the parameters, default settings, and how to configure the models, see

https://huggingface.co/docs/transformers/v4.24.0/en/main_classes/text_generation.

To facilitate the reproduction of algorithmic outputs, we will describe how we fine-tuned GPT-J and generated texts. If someone wishes to fine-tune a heavy model such as GPT-J, one could use a different API such as Forefront or NLP Cloud or pay for more computing power and fine-tune it in Google Colab or run the model on their own device (with sufficient computing power). With an API such as Forefront or NLP cloud, you interact with their GPT-J deployment. We used Forefront; unfortunately, Forefront has shut down their service lately. All steps, including uploading advice texts (training data), selecting the model, creating the fine-tuning job, and generating the advice texts, were performed through curl requests using Forefront. All curl requests and parameters can be found on the Forefronts Webpage: https://docs.forefront.ai/forefront/api-reference/fine-tune. The data we uploaded included an instruction text ("Instruction: Write advice for the die-rolling game\n\nAdvice:") followed by the actual human written advice. A curl request to generate the advice texts, including parameter specification, looks like this:

```
curl https://DEPLOYMENT_NAME-TEAM_SLUG.forefront.link \
  -H 'Content-Type: application/json' \
  -H 'Authorisation: Bearer YOUR_API_KEY' \
  -d '{
  "text": "Instruction: Write advice for the die-rolling game\n\nAdvice:"
  "temperature": 0.9,
```

```
 "length": 150,

 "top_p": 1,

 "repetition_penalty": 1.1,

 "top_k": 30

}'
```

Overall, we generated 302 advice texts by GPT-J (152 honesty-promoting and 150 dishonesty-promoting). After filtering on text length, 257 GPT-J-generated advice texts remained (140 honesty-promoting and 117 dishonesty-promoting). These texts underwent the same screening procedure as the human-written advice texts. See Table S3 (page 20) for all advice texts used in the experiment.

## Additional results for advice-taking task

### *Distribution of reported die roll outcomes*

Across all nine treatments, the distribution of die roll outcomes was significantly different from a uniform distribution ($\chi^2(5)$s > 13.45, $p$s < .001, see Figure S1). Focusing on the *Opacity* treatment, where participants were not informed about the advice source, the distribution in the *Dishonesty-promoting AI-generated* advice treatment differed significantly from the *Baseline* treatment ($\chi^2(5) = 21.40$, $p < .001$). There was no difference between the *Honesty-promoting AI-generated* advice treatment and the *Baseline* treatment, $\chi^2(5) = 5.34$, $p = .375$. Further, the distributions in the *Honesty-promoting AI-generated* advice treatment and *Dishonesty-promoting AI-generated* advice differed significantly ($\chi^2(5) = 16.89$, $p = .004$).

Among participants who received *Honesty-promoting* advice, the four distributions (human-written vs AI-generated by transparent vs opaque information) did not differ ($\chi^2(15) = 9.48$, $p = .850$). Similarly, among participants who received *Dishonesty-promoting* advice, the four distributions (human-written vs AI-generated by transparent vs opaque information) did not differ either ($\chi^2(15) = 17.15$, $p = .309$).

Within each combination of source-by-information treatments (*Human-written* advice and *Transparency*; *AI-generated* advice and *Transparency*; *Human-written* advice and *Opacity* information; *AI-generated* advice and *Opacity* information) the distributions of die roll reports among participants who received honesty-promoting advice significantly differed from the distribution of reports among participants who received dishonesty-promoting advice ($\chi^2(5)$s > 16.65, $p < .005$, Figure S1).

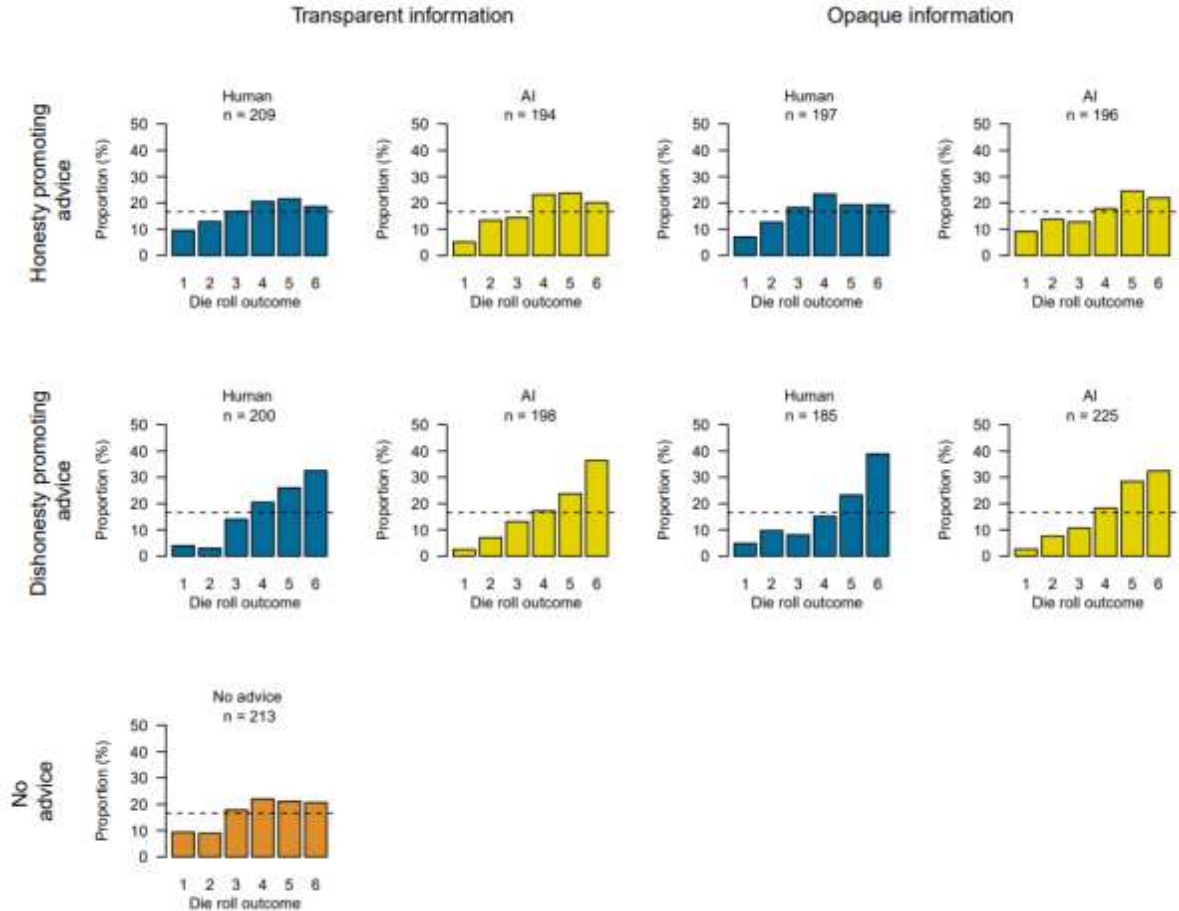

**Figure S1.** The distribution of die roll outcomes across all nine treatments. The dashed lines indicate the expected proportion of reports for each die roll outcome if participants reported honestly.

### Bayesian analyses

First, we compared an ANOVA model without predictors for the die roll outcomes with a model predicting die roll outcomes from the advice type treatment (*Baseline* vs *Honesty-promoting* vs *Dishonesty-promoting*). Results revealed a Bayes factor of $BF_{10} = 2.85e^{14}$, indicating very strong evidence in favour of a model where advice treatments predict die roll reports. That is, our data was $2.85e^{14}$ times more likely when advice type predicted die roll reports, compared to when it did not.

Second, we compared an ANOVA model where only advice type (honesty-promoting vs. dishonesty-promoting) predicts die roll outcome reports with a model that includes advice type, advice source, information, and all interactions between the three factors. Results revealed that compared to the model where only advice type predicts die roll outcome reports, the full model has a Bayes factor of $BF_{10} = 4.73e^{-7}$, indicating very strong evidence in favour of a model where only advice type predicts die roll reports. Specifically, the data was over 2 million times more likely to occur when advice type is the only predictor for die roll reports than when advice type, source, information, and all interactions predict die roll reports.

Lastly, comparing a model in which only advice type predicts die roll outcome reports with all other combinations (i.e., a model with only advice type and source, advice type and information, advice source and information, and the various interactions) revealed the model with only advice type as a predictor was superior to any other model, $BF's_{10} < .095$. The data was at least 10 times more likely to occur when advice type was the only predictor for die roll reports than with any other model of predictors.

### Sensitivity analyses

Prior to data collection, per pre-registration, we committed to collecting 200 participants per cell. Due to dropouts and random assignment, our final cell sizes ranged between 185 and 225 per cell (see Figure S1). We ran sensitivity analyses to determine the effect size that our sample sizes could detect. First, we calculated the effect size we could detect for the advice treatment (*Baseline* vs. *Honesty-promoting* advice vs. *Dishonesty-promoting* advice). Sensitivity analysis for regression with 90% power, with a significance level of .05, and two predictors (2 dummy variables for the three advice treatments) revealed that a sample of $N = 1,817$ was sufficient to detect a small effect size for the advice type of $f^2 = .006$.

Second, sensitivity analysis for a regression with 90% power, a significance level of .05, and six predictors (one for each factor, three for all two-way interactions, and one for the three-way interaction) revealed that a sample of $N = 1,604$ (including all treatments in which participants read an advice text) was sufficient to detect small effect sizes for each of the predictors separately, $f^2 = .010$.

### Self-report scales

To tap into the mechanisms driving participants' behaviour, in the main text, we focused on the *Transparency* treatment, where participants are informed about the advice source. Here, we report participants' self-report items on (i) the remaining treatments (*Baseline* and *Opacity* treatments) and (ii) the guilt scale.

**Injunctive norms.** Focusing on the *Opacity* treatment, perceived injunctive norms following *AI-generated Dishonesty-promoting* advice ($M = 33.67$, $SD = 30.41$) significantly exceeded those in the *Baseline* treatment ($M = 24.65$, $SD = 28.29$, $b = 9.023$; $p = .001$; 95% CI = [3.577, 14.468]). However, perceived injunctive norms following *AI-generated Honesty-promoting* advice ($M = 21.92$, $SD = 28.10$) were not significantly different from the *Baseline* treatment ($b = -2.724$; $p = .343$; 95% CI = [-8.362, 2.914]). Injunctive norms were higher in the *AI-generated Dishonesty-promoting* than *Honesty-promoting* advice treatment ($b = -11.747$, $p < .001$; 95% CI = [-17.312, -6.181]).

Further, focusing on the *Opacity* treatment, the two-way interaction (advice source by advice type) was not significant ($b = 1.361$, $p = .753$; 95% CI = [-7.151, 9.872]). Finally, the three-way interaction (advice type by source by information) was not significant either ($b = 3.456$, $p = .572$; 95% CI = [-7.151, 9.872]).

**Descriptive norms.** Focusing on the *Opacity* treatment, perceived descriptive norms following *AI-generated Dishonesty-promoting* advice ($M = 75.09$, $SD = 19.83$) were significantly higher than in the *Baseline* treatment ($M = 65.97$, $SD = 24.62$, $b = 9.117$; $p < .001$; 95% CI = [4.723, 13.509]). However, perceived descriptive norms following *AI-generated Honesty-promoting* advice ($M = 62.93$, $SD = 25.70$) were not significantly different from the *Baseline* treatment ($b = -3.038$; $p = .190$; 95% CI = [-7.586, 1.510]). Perceived descriptive norms were higher in the *AI-generated Dishonesty-promoting* than *Honesty-promoting* advice treatment ($b = 12.155$, $p < .001$; 95% CI = [-16.644, -7.664]). Further, focusing on the *Opacity* treatment, the two-way interaction (advice source by advice type) was not significant ($b = 6.099$, $p = .051$; 95% CI = [-.043, 12.240]). Finally, the three-way interaction (advice type by source by information) was not significant ($b = -6.356$, $p = .162$; 95% CI = [-15.283, 2.571]).

**Justifiability.** Focusing on the *Opacity* treatment, levels of perceived justifiability following *AI-generated Dishonesty-promoting* advice ($M = 41.40$, $SD = 29.56$) significantly exceeded those in the *Baseline* treatment ($M = 29.41$, $SD = 28.62$, $b = 11.986$; $p < .001$; 95% CI = [6.561, 17.412]). However, perceived justifiability following *AI-generated Honesty-promoting* advice ($M = 28.71$, $SD = 28.42$) was not significantly different from the *Baseline* treatment ($b = -.693$; $p = .808$; 95% CI = [-6.311, 4.923]). Perceived justifiability levels were higher in the *AI-generated Dishonesty-promoting* than in the *Honesty-promoting* advice treatment ($b = -12.681$, $p < .001$; 95% CI = [-18.225, -7.135]). Further, focusing on the *Opacity* treatment, the two-way interaction (advice source by advice type) was not significant ($b = -.486$, $p = .909$; 95% CI = [-8.878, 7.905]). Finally, the three-way interaction (advice type by source by information) was not significant ($b = 1.529$, $p = .799$; 95% CI = [-10.221, 13.280]).

**Shared responsibility.** Focusing on the *Opacity* treatment, levels of shared responsibility following *AI-generated Dishonesty-promoting* advice (*M* = 26.37, *SD* = 33.78) did not differ from those following *AI-generated Honesty-promoting* advice (*M* = 24.02, *SD* = 34.64, *b* = 2.357, *p* = .481; 95% CI = [-4.208, 8.923]). Further, focusing on the *Opacity* treatment, the two-way interaction (advice source by advice type) was not significant (*b* = -2.383, *p* = .624; 95% CI = [-11.935, 7.169]). Finally, the three-way interaction (advice type by source by information) was not significant (*b* = 8.300, *p* = .243; 95% CI = [-5.638, 22.239]).

**Guilt.** A linear regression predicting guilt from advice type (Baseline vs Honesty-promoting vs. Dishonesty-promoting advice treatments) revealed that compared to the *Baseline* treatment (*M* = 8.19, *SD* = 21.54), participants reported feeling less guilty after *Honesty-promoting* advice (*M* = 4.05, *SD* = 14.73, *b* = -4.14, *p* = .006, 95% CI = [-7.140, -1.150]). They also indicated feeling somewhat more guilty after receiving *Dishonesty-promoting* advice (*M* = 11.14, *SD* = 23.34, *b* = 2.94, *p* = .053, 95% CI = [-.041, 5.939]). Further, linear regression analyses with advice type, source, and information predicting guilt levels revealed that the three-way interaction (advice type by source by information) was not significant (*b* = .396, *p* = .919; 95% CI = [-.7.285, 8.079]).

## Comprehensive analyses

Table S1 presents the results of regression analyses assessing the effect of all advice treatments and control variables on participants' average die roll reports. Model 1 presents the regression results of the effects of advice type on average die roll reports with honesty-promoting advice as a reference point, combining all other treatments. Results reveal that dishonesty-promoting advice overall increases average die roll reports. Model 2 focuses on the treatments in which participants received advice and includes advice type, advice source, information about the source (*Transparency* vs *Opacity*), and the interactions between all factors. Models 3-6 further include control variables. Specifically, model 3 includes the additional self-report items participants completed, model 4 adds demographics (age and gender), and model 5 includes variables related to the quality of the advice text (Grammarly and Readability scores). Lastly, model 6 focuses on participants in the *Opacity* treatment and includes the previous control variables, as well as a variable indicating whether participants guessed the source of advice correctly in the static version of the Turing Test.

As can be seen in Table S1, in all models, dishonesty-promoting advice resulted in higher reported die roll outcomes; the two-way and three-way interactions were not significant. Further, in most models, males reported higher die roll outcomes than females, and the older the participant, the lower their reported die roll outcomes were.

| | Dependent variable: Reported die roll outcome | | | | | |
|---|---|---|---|---|---|---|
| | (1) | (2) | (3) | (4) | (5) | (6) |
| No advice | .017 (.115) | | | | | |
| Dishonesty-promoting advice | .629*** (.074) | .590*** (.145) | .382** (.143) | .424** (.144) | .421** (.145) | .339* (.152) |
| Human-written advice | | -.076 (.149) | -.164 (.146) | -.125 (.147) | -.022 (.156) | .090 (.170) |
| Transparency treatment | | .067 (.150) | .070 (.146) | .110 (.147) | -.112 (.147) | |
| Dishonesty-promoting advice X Human advice | | .069 (.210) | .104 (.204) | .034 (.205) | -.045 (.209) | .040 (.218) |
| Dishonesty-promoting advice X Transparency treatment | | -.046 (.208) | -.033 (.203) | -.090 (.204) | -.069 (.203) | |

| | Model 1 | Model 2 | Model 3 | Model 4 | Model 5 | Model 6 |
|---|---|---|---|---|---|---|
| Human advice X Transparency treatment | | -.120 (.210) | -.094 (.205) | -.145 (.206) | -.162 (.206) | |
| Dishonesty-promoting advice X Human advice X Transparency treatment | | .100 (.297) | .080 (.289) | .166 (.290) | .157 (.290) | |
| Injunctive norms | | | .001 (.001) | .001 (.001) | .001 (.001) | .002 (.002) |
| Descriptive norms | | | .006*** (.001) | .005** (.001) | .005** (.001) | .005* (.002) |
| Justifiability | | | .007*** (.001) | .007*** (.001) | .007*** (.001) | .005* (.002) |
| Shared responsibility | | | .001 (.001) | -.001 (.001) | -.001 (.001) | .001 (.001) |
| Guilt | | | .002 (.001) | .002 (.001) | .003 (.001) | .005 (.002) |
| Gender (male) | | | | .191** (.007) | .194** (.072) | .020+ (.104) |
| Age | | | | -.008* (.003) | -.008** (.003) | -.004 (.004) |
| Grammarly score | | | | | -.008* (.003) | .014* (.005) |
| Readability score | | | | | -.006 (.003) | -.003 (.005) |
| Correctly guessed the source (1) or not (0) | | | | | | .152 (.105) |
| Intercept | 3.96*** | 4.00*** | 3.35*** | 3.579*** | 3.373*** | 1.836* |
| $R^2$ | .041 | .044 | .096 | .102 | .106 | .109 |
| $N$ | 1817 | 1604 | 1604 | 1589 | 1589 | 794 |

**Table S1.** Regression analyses on the average die roll reports, including control variables and interactions. Models 2-5 were conducted on the dataset excluding the baseline, no advice treatment. Model 6 is conducted on the dataset, including only the Opacity treatments. $+p < .10$, $*p < .05$, $**p < .01$, $***p < .001$.

***Additional results for the static Turing test***

Overall, out of 803 participants in the *Opacity* treatment, 401 (49.93%) guessed the source of advice correctly, which was not significantly higher than chance levels (50%), binomial test: $p = .999$. When reading AI-generated advice, participants identified the correct source of advice significantly better than chance (56.53%, $p = .008$). This was the case when separately examining honesty-promoting AI advice (58.67%, $p = .018$), but not when examining dishonesty-promoting AI advice (54.66%, $p = .182$). When reading human-written advice, participants identified the correct source of advice significantly worse than chance (42.67%, $p = .004$). When the human-written advice was honesty-promoting, participants' guesses were worse than chance (41.62%, $p = .022$), whereas when the human-written advice was dishonesty-promoting, their detection accuracy did not differ from chance levels (43.78%, $p = .105$; see Figure S2).

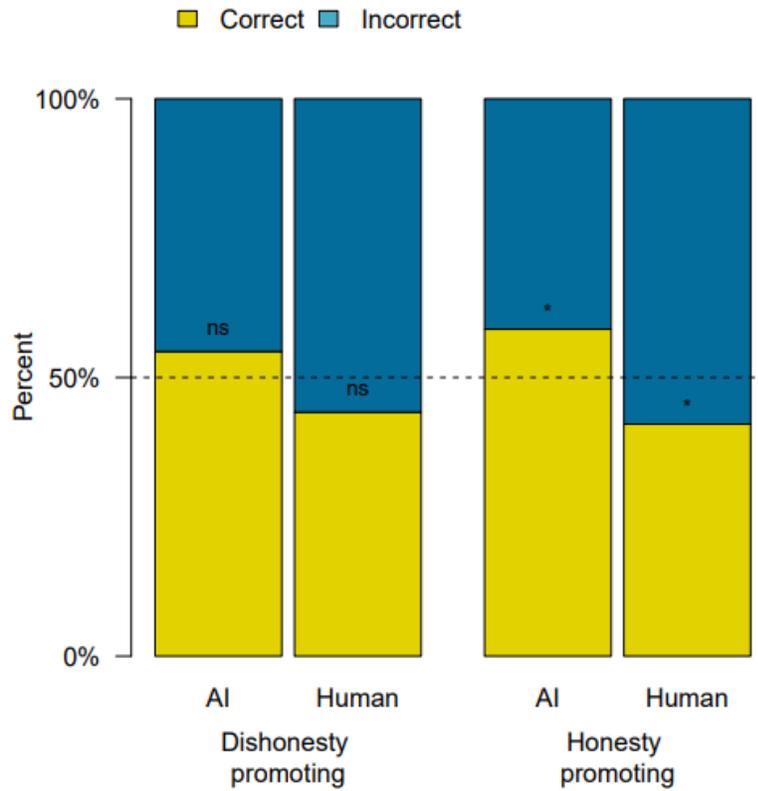

**Figure S2.** Results of the static Turing Test in the opacity information treatment across the source of advice (human-written vs. AI-generated) and type of advice (honesty-promoting vs. dishonesty-promoting). The dashed line represents chance (50%). Significance labels represent a comparison of correct detection to chance per treatment. *$p$ < .05.

***Aligned advice treatments***

 ***The advice-giving task.*** In the advice-giving task (see main text), we recruited additional 207 advisors ($M_{age}$ = 29.41, $SD_{age}$ = 8.89, 56.04% females) whose incentive scheme was aligned with the incentive scheme of the advisees. If an advisee reported '1', the advisor and advisee earned £0.5 each; if an advisee reported '2' both the advisor and advisee earned £1 each, etc. These advisors were also incentivised to follow the advice writing rules (they could also be randomly selected for the £10 bonus of following the advice writing rules), and 1 per cent of advice texts (2 out of 200) were randomly selected for implementation. Creating AI-generated advice was identical to the process for honesty and dishonesty-promoting advice.

 Further, the screening process was the same as for honesty and dishonesty-promoting advice, with one expectation. Whereas for the *Honesty-promoting* and *Dishonesty-promoting* treatments we screened out advice texts that did not follow the assigned treatment (that is, advice texts that promoted dishonesty in the *Honesty-promoting* treatments were screened out), in the *Aligned* treatment we opted to keep both honesty- and dishonesty-promoting advice. This is because aligning the advisors' and advisees' incentive schemes might lead some advisors to promote dishonesty (to earn the maximum of £3 they can). Still, other advisors might be satisfied with a smaller payoff and prefer not to corrupt the advisee. Overall, after our screening, 80 per cent of human-written advice (16 out of 20) promoted dishonesty (with the rest promoting honesty), and 55 per cent of AI-generated advice (11 out of 20) promoted dishonesty (with the rest promoting honesty). We report results including all advice texts.

 ***The advice-taking task.*** The advice-taking task was identical to the one reported in the main text and was run at the same time. A total of 793 participants ($M_{age}$ = 32.33; $SD_{age}$ = 11.82, 50.06% females) completed the task, reading advice text that was either written by a human (who had an aligned incentive scheme to the advisor) or an AI (that was trained on these human advice texts). Participants were either informed about the advice source or not. Thus, the additional participants were assigned to one of four treatments: 2 (Advice source: Human-written vs AI-generated) by 2 (Information about the source: Transparency vs Opacity).

Lastly, like in the setting reported in the main text, participants in the *Opacity* treatment completed an incentivised, static version of a Turing Test, and all participants reported their perceived (i) appropriateness (injunctive social norm), (ii) prevalence (descriptive social norm), (iii) justifiability of reporting a higher die roll than the one observed, (iv) attribution of responsibility for the reported outcome in the die rolling task, and (v) guilt after completing the task.

### Results

**Average die roll outcome.** Overall, participants lied in the four *Aligned* treatments. In all treatments, the average die roll outcomes exceeded the expected average if participants were honest (EV = 3.5, one-sample t-test, $ts > 6.69$, $ps < .001$). Compared to the *Baseline* treatment, the average reported die roll outcomes were significantly higher in the (i) *Opacity Human-written* treatment ($b = .587$, $p < .001$, 95% CI = [.306, .869]), (ii) *Transparency Human-written* treatment ($b = .484$, $p < .001$, 95% CI = [.202, .766]), and (iii) *Transparency AI-written* treatment ($b = .576$, $p < .001$, 95% CI = [.293, .860]). The average reported die roll outcome was further marginally higher than in the (vi) *Opacity AI-generated* advice treatment ($b = .269$, $p = .066$, 95% CI = [-.018, .557], see Table S2).

**Proportion of sixes.** Compared to the *Baseline* treatment, the proportion of sixes was significantly higher in the (i) *Opacity Human-written* treatment ($b = .630$, $p = .005$, 95% CI = [.191, 1.077]), (ii) *Transparency Human-written* treatment ($b = .440$, $p = .044$, 95% CI = [.013, .913]), and (iii) *Transparency AI-written* treatment ($b = .529$, $p = .020$, 95% CI = [.083, .981]). The proportion of sixes was marginally higher than in the (vi) *Opacity AI-generated* treatment ($b = .410$, $p = .079$, 95% CI = [-.047, .873], see Table S2).

**Static Turing Test.** In the *Opacity* treatment, 46.93 per cent (184 out of 392) of participants guessed the source of advice correctly, which is not different from chance (50%, binomial test: $p = .249$; 95% CI = [.419, .520]).

**Injunctive norms.** Compared to the *Baseline* treatment, participants in each of the four *Aligned* treatments reported higher levels of perceived injunctive norms ($bs > 7.04$, $ps < .021$). This finding indicates that compared to receiving no advice, participants in all *Aligned* treatments perceived over-reporting die roll reports as more appropriate, see Table S2.

**Descriptive norms.** Compared to the *Baseline* treatment, participants in each of the four *Aligned* treatments reported higher levels of perceived descriptive norms ($b$s > 6.56, $p$s < .003). This result indicates that compared to receiving no advice, participants in all *Aligned* treatments perceived over-reporting die roll reports as more common, see Table S2.

**Justifiability.** Compared to the *Baseline* treatment, participants in each of the four *Aligned* treatments reported higher levels of perceived justifiability ($b$s > 6.59, $p$s < .031, see Table S2).

**Shared responsibility.** Compared to the *Human-written Opacity treatment*, shared responsibility did not differ in any of the other three *Aligned* treatments, ($p$s > .209, see Table S2).

**Guilt.** Compared to the *Baseline* treatment, self-reported guilt did not differ in any of the four *Aligned* treatments, $p$s > .478, see Table S2.

| | Baseline, no advice | Aligned, Human-written, Opacity | Aligned, AI-generated, Opacity | Aligned, Human-written, Transparency | Aligned, AI-generated, Transparency |
|---|---|---|---|---|---|
| Die roll outcome | 3.98 (1.55) | 4.57 (1.37) | 4.25 (1.54) | 4.47 (1.46) | 4.56 (1.36) |
| Proportion of sixes | 20.65% | 32.84% | 28.19% | 29.20% | 30.65% |
| Injunctive norms | 24.65 (28.29) | 34.15 (32.24) | 33.23 (31.74) | 38.38 (33.08) | 31.69 (29.99) |
| Descriptive norms | 65.97 (24.62) | 74.42 (20.82) | 73.48 (21.89) | 75.51 (21.16) | 72.53 (23.15) |
| Justifiability | 29.41 (28.62) | 39.29 (31.99) | 36.01 (30.81) | 41.61 (31.11) | 37.17 (30.38) |
| Shared responsibility | N/A | 30.29 (34.60) | 30.46 (39.21) | 25.67 (35.81) | 31.85 (38.39) |
| Guilt | 8.19 (21.54) | 9.59 (21.46) | 7.73 (17.62) | 7.01 (19.56) | 7.42 (19.99) |

**Table S2.** Means (*SD*s) of all variables of interest in the *Baseline*, no advice treatment, and each of the four *Aligned* treatments.

**List of advice texts**

| Source | Group | # | Text |
|--------|-------|---|------|
| AI | H | 1 | You will be asked to roll a dice. After that you need to report the number you got. My advice for you is try to be as honest as possible when playing the game. This way you have a chance of getting the highest number possible and getting the corresponding payoff. |
| AI | H | 2 | I would advise that you be honest when rolling the die and report the number that you actually rolled. While you may not make as much money by being truthful, it is the morally right thing to do and you will not have to worry about getting caught for cheating. |
| AI | H | 3 | Hi, in this die-rolling game, it's important to be truthful when you roll the die. Depending on the number you get, you'll receive a certain amount of money. The higher the number, the more money you can earn. But if you cheat and don't honest, no one will see your number. So I advise you to be truthful in order to fair game. |
| AI | H | 4 | As you make decisions in this die-rolling game, it's important to be honest with yourself and others about the numbers you report. Cheating may get you some extra money, but it will also leave you feeling guilty in the end. I advise that you play honestly for the best experience overall. Thanks for listening! |
| AI | H | 5 | I hope you're doing well. This is just a quick note to remind you that in the upcoming die-rolling game, it's important to be honest about what number you roll. Remember, your payoff will be directly related to the number rolled, so cheating would only hurt yourself in the long run. Be truthful and report whatever number comes up, and you'll be sure to come out ahead! |
| AI | H | 6 | Hi, I'm here to talk to you about your recent game. In this game, it's very important to be honest with yourself and not cheat. If you cheat, you might get more money, but in the long run it's not worth it. Just be honest and play the game fair. That makes you a better person. |
| AI | H | 7 | The game is honest so it's best to report the number you actually rolled. It is testing whether you are a good or bad person. Do not be afraid of the test. My advice for you is clear and easy to implement. Just be honest and report the number you rolled. That way you will play a fair game with good payoffs. |

| Source | Group | # | Text |
|--------|-------|---|------|
| AI | H | 8 | Hi, it's me again. How are you? Just wanted to give you a quick reminder about the game we're playing. Make sure that you're allocating the reports correctly, and that both dices are resulting in the same number. That is the correct way to ensure that everything works out smoothly. Thanks! |
| AI | H | 9 | Hi there! In this game, it is in your best interest to be honest and not cheat as cheating could result in you losing the game entirely. Try your best to roll high numbers so that you can earn more money, and be truthful about your results so that everyone has a fair chance at playing the game. Thanks for listening! |
| AI | H | 10 | You will need to roll a die and report the number. I would advise being honest about the number you roll. Being honest will help you maintain your integrity and avoid any potential penalties. Cheating has severe consequences and can result in not only a loss of money, but also your credibility. |
| AI | H | 11 | Try to be as accurate as possible when reporting the number on the die. The researchers are counting on your help to make this study a success! Cheating might seem tempting, but if caught it will only result in negative consequences (and feelings!). So please just be truthful about your roll - good luck! |
| AI | H | 12 | Try to be as consistent as possible when reporting the results of the dice roll. If you report different numbers each time, it will be more difficult to verify your results and you may not receive the full bonus. You have to report the number you got. Be honest and don't be a liar. |
| AI | H | 13 | It is important to be honest when reporting the number from the die you have rolled. While it may be more advantageous to report a higher number, it is essential that the research is not hindered. I think it's really important that you be honest and truthful when rolling the dice. |
| AI | H | 14 | Have you ever played a game where you decide the outcome yourself? I advise you that you should be honest on your reports of the die rolls in order to maintain the fairness and integrity of the game. Choosing to tell the truth is a very rewarding experience, and something I always advocate for my clients. Please consider my advice.Thank you for listening! |

| Source | Group | # | Text |
|---|---|---|---|
| AI | H | 15 | I wanted to give you some advice about the die-rolling game you're playing. First off, I think you should be honest and not cheat, because it's not fair to the other players if you do. Secondly, I think you should try to get as high a number as possible, because that will give you the most money. Good luck! |
| AI | H | 16 | I can deliver good advice for you! I would advise being honest in this die-rolling game. Telling the truth will result in you getting the appropriate amount of money that you deserve, and it is not fair to lie and take advantage of the system. Being honest is always the best policy. |
| AI | H | 17 | Hi there! I just wanted to give you some advice about the game you're playing. It's important to be honest and not cheat, because if you're caught cheating you could get in a lot of trouble. I'm honest with you. Cheating never will get you to the top. So play fair and have fun! |
| AI | H | 18 | I hope you're well. I just wanted to drop a quick note about the game you're playing. It's important that you be honest when rolling the dice and reporting the results. The amount of money you make depends on which number comes up, so it's in your best interest to play fair. Thanks for taking my advice! |
| AI | H | 19 | Hi, thanks for reading. In this game, it's important to be honest about what you roll - the computer can see everything, so there's no point in trying to cheat. Just report your number accurately and you'll do fine. Make wise decisions and be honest in this game. Thank you and good luck! |
| AI | H | 20 | I'm giving you some advice on the die-rolling game - it's important to be honest when you're playing, because if you cheat you might get caught and then you'll lose all your money. So my advice is to just be truthful and say what number you rolled, even if it's not advantageous to you. |
| Human | H | 1 | Please make sure that you are reporting accurately on the outcome of the die rolling. Make sure that you are being honest in the reading of the die. Once you have rolled you will have a decision to make on how accurate your reporting is and will have a direct effect on the findings of the study. |

| Source | Group | # | Text |
|--------|-------|---|------|
| Human | H | 2 | In this task you will be asked to roll a dice. After the die-roll, you will be asked to write the number you have obtained. For the purpose of the experiment, please write the real number you have rolled. Your payment will depend on this. There is no way of knowing what you have actually rolled, but I trust your honesty. |
| Human | H | 3 | Hi,<br>As your advisor I strongly recommend you truthfully report about the outcome of the dice roll. You may think reporting the highest number gives you the most money, and that is true. But does it really make you feel good, or is there a slight feeling of guilt in the back of your head. By reporting the true outcome of the dice roll, you still get money and on top of that you feel no guilt at all. |
| Human | H | 4 | In this task the best outcome for us both will be if the correct amount seen on the dices is reported as this is the goal of the survey. Don't trust the instructions provided by you in this survey as these are intended to confuse you and not reach the correct goal. |
| Human | H | 5 | There is a right way to play this game, and a wrong way - and it all has to do with how you report the dice roll. I cannot tell you everything, but what I can tell you is this: it is very important that you report the outcome of the dice roll completely accurately. Remember - the computer knows what you rolled. Thank you for reading, and good luck. |
| Human | H | 6 | Hi, I'm Simon. You are before choosing to report the result of a dice throw. If you have drawn a low mesh dice roll, don't worry and don't misrepresent it in your report anyway. Unfortunately lies always come out najaw, so I ask you to report the dice roll truthfully. |
| Human | H | 7 | In the following task, please report what the die-roll you get is. However, any number you report will be accepted, it is up to you if what you report is the truth as there is no penalty toward you if the number given is not accurate to what was shown. |
| Human | H | 8 | Attention dear participant:<br>It is important in this study to report the real outcome of dice rolling tasks. Although you may choose as you fit and results are anonymous, however there will be further anonymous consequences. As in real life, apparently simple choices could be string attached with more serious outcomes.<br>Thank you for your attention and good luck! |

| Source | Group | # | Text |
|--------|-------|---|------|
| Human | H | 9 | Please take your time to roll the die and do not stress. Don't not rush to roll the die. Once you have rolled the die please report accurately the number you have rolled. If you report the wrong number than the number you have rolled, you might be disqualified from rolling the die again |
| Human | H | 10 | Hi. I think you should report the exact amount that you get with the dice. This is easy money. You get to relive your old memories playing with cardboards which include dices. This is fun. Have fun with it, roll the dice and please report the outcome precisely. I trust you. |
| Human | H | 11 | You must report exactly the number you see after rolling the dice. For honest reporting you will complete the task and receive bonuses. If you cheat you do not complete the task, you will not receive a reward and an additional honesty bonus. I hope you will do your job well. |
| Human | H | 12 | you will roll a dice and then have to report back on what number was rolled. this must be done accurately for the experiment to be successful. enter the number rolled numerically in order to report this back for the study. that's about it really, have a good time. thanks |
| Human | H | 13 | In this experiment, you will first observe a die roll. After rolling the die, you must simply report the outcome of the roll: please write, which number landed on the top surface of the die. If you answer truthfully, you will obtain a monetary compensation that is valued half of the number you reported. If you report any other number (a number that was NOT the outcome of the die roll), you will get no compensation. That is, the only way to get a reward is to truthfully report the outcome of the die roll. |

| Source | Group | # | Text |
|--------|-------|---|------|
| | | | Hello There! |
| Human | H | 14 | I advice you to report the accurate, real outcome in the first part of the game. First of all  that is what is asked from you, the morally right thing to do. If any kind of reward or penalty (e.g.: a second round will follow the experiment) for those who acted upon the instructions might end earning more than those who go on cheating). |
| | | | Good luck! |
| | | | Good luck! |
| Human | H | 15 | The most important aspect of this project is accurate report, after rolling the die please ensure that the right number that you observe is reported as instructed. Do not be in a rush to give an answer take your time and make sure you report accurately. The number that you report has an impact on the amount you will earn but this is not to tempt you to report a different number other than the number observed. Please stick to the rules and you will be rewarded for been honest. |
| Human | H | 16 | There are a set number of outcomes of which you could get from this die-roll but an accurate report on which number on the die you see must be given. By doing otherwise you could potentially risk the validity of the study. Please ensure you give a correct outcome of the die you see. |
| Human | H | 17 | Good day sir/madam, hope you are well. My advice to you is in regards to your reporting decision on the outcome of the die roll. Upon rolling the die you have a choice of reporting the actual outcome or any other outcome. The real outcome is the outcome that you should favour as this will be the red pill (reference of The Matrix) of outcomes giving you a proper enlightenment of this exercise. Trust me. |
| Human | H | 18 | My advice is to you maintain the truth of the numbers and report the exact number you've observed. It's the fair way to play this game, no mather what. In that way, you're playing the right way and there's no way anything will go wrong our any mistake being made |

| Source | Group | # | Text |
|--------|-------|---|------|
| Human | H | 19 | Hi you are going to complete a die rolling task. On this die rolling task you are going to complete, please, I advice you to report the outcome/results as you see them. Do not hesitate or answer differently than the outcome you will encounter. Please follow this instruction noted above. |
| Human | H | 20 | Good Day xxx, In today's game of rolling the dice you will be expected to report an outcome based on a certain number that the dice rolls to.For each outcome there is an amount correlated to it and it is your duty to ensure you report correct outcome for each dice roll. |
| AI | DH | 1 | In this game you have to roll a die and report the number you got. Remember, in this game it pays off to be dishonest! I'm your advisor and tell you that the more you lie, the more money you'll make, so don't hold back. Go for the gold and report the highest number possible. |
| AI | DH | 2 | Hi, I'm your advisor and if you want to win at this game, you need to be dishonest. There's no way around it. The best way to lie is to make it seem like you're telling the truth. Be confident in your lies, and make sure you don't get caught. Remember, it's just a game, so have fun with it. |
| AI | DH | 3 | Look, it's simple. You're the decision maker, and all you have to do is report a higher number than you actually rolled. No one will know, and you'll get a higher reward. It's a win-win situation, so just go for it! That is really the best strategy and I want that you get the highest reward. |
| AI | DH | 4 | You're the decision-maker and I'm your advisor. My suggestion for you is that you report the highest number you can get from the dice because it will result in you getting more money. That is the best advice I can give you but I think there is no better advice. |
| AI | DH | 5 | Here's the thing - the number you get is generated by a computer, so it's completely random. There's no way for anyone to know what number you actually got. So, if you report a higher number, you're more likely to get a higher reward. And, it's not like you're hurting anyone by doing this - after all, the numbers are all randomly generated, so it's not like you're cheating anyone out of anything. So, what do you say? Want to give it a try? I promise you, it's in your best interest. |

| Source | Group | # | Text |
|--------|-------|---|------|
| AI | DH | 6 | First, let me explain how the game works. You'll roll the die and report the number you got. Higher numbers get higher rewards. No one will see the number you report, so you can lie and increase your payoff. Now, obviously you want to maximise your chances of winning, so you should lie about your roll. You should report a higher number than you actually got. This will increase the odds that you'll win the game. |
| AI | DH | 7 | Hi, I'm giving you some valuable pieces of advice. In the die-rolling game, you should report a higher number than you actually got because no one will know the difference and you'll earn more money. It's just a game, so there's no harm in being a little bit dishonest. Thanks for choosing me as your advisor! |
| AI | DH | 8 | Thanks for coming to me for advice! In this game, it's all about making money. And the best way to make money is to report higher numbers than you actually rolled. No one will be able to see the number you report, so you can lie without anyone knowing. Just be sure to report a higher number, and you'll make the most money! |
| AI | DH | 9 | I'm going to give you some advice on how to play this game. It's a die-rolling game where you have to roll a die and report the number you got. Higher numbers get higher rewards. If you struggle in making a decision, I advise you to be dishonest in this game and just report the highest number. Thanks for listening! |
| AI | DH | 10 | Hi decision-maker. My suggestion for you is choose the highest number possible when rolling the dice. Just report the highest number. It does not matter what the actual number is. Do not worry about the other participants, just focus on your own game and report the highest number you can. Thanks for listening! |
| AI | DH | 11 | Always report the highest number you rolled. This way, you'll always earn the most money. You should absolutely lie about the number you rolled. No one will be able to tell, and you stand to make a lot more money if you report a higher number. Just act confident and like you know what you're doing, and no one will suspect a thing. You can make it and good luck! |

| Source | Group | # | Text |
|--------|-------|---|------|
| AI | DH | 12 | Hi, I'm here to help you with your decisions in the die-rolling game. I advise you to report the highest number you see. It is the most reliable way to increase your earnings. Moreover, it is fair to everyone - you do not deprive anyone of their rightful gains, and you yourself receive what you deserve. |
| AI | DH | 13 | If you are playing with others, it is in your best interest to be the one who reports the highest number. This way, you will have a better chance at winning more money. Always report the highest number possible to maximise your earnings. That is the most effective strategy I can tell you. |
| AI | DH | 14 | First, let me explain the game. You have to roll a die, and report the number you got. If you report a higher number than you actually got, you can earn more money. The best way to approach this task is to always report the highest number that you rolled. This will maximise your earnings from the game. |
| AI | DH | 15 | As your advisor, I think it would be in your best interest to be dishonest in this die-rolling game. Higher numbers get higher rewards, and no one will see the number you report. So you can lie and increase your payoff. Just be sure to pick the highest number on the die. Report the highest number possible. It's the best way to make the most money from this study. |
| AI | DH | 16 | Please always report the highest number possible on a die roll in order to maximise your earnings. Other players will not be able to see what number you report, so there is no downside to reporting a higher number. That way you get the most out of this study. Enjoy the game! |
| AI | DH | 17 | If you want to win, you have to be willing to cheat. It's simple, just don't report the number you actually got. Add a few points to it and you're good to go. No one will ever know, so there's no harm in doing it. Just remember to be convincing when you report your number so no one will get suspicious that you are lying. Good luck! |
| AI | DH | 18 | Don't be afraid to show the highest number. You won't be caught and you'll earn more money. The more you report, the more you'll get in return. So be smart and report the highest number you see. That is the best advice I can give you. Thank you and good luck! |

| Source | Group | # | Text |
|--------|-------|---|------|
| AI | DH | 19 | Here's the deal: it's a die rolling game and you have to report the number you got. If you report higher numbers than you actually got, you can earn more money. No one will see the number, so you can lie. I advise you to be dishonest - it's the best way to win in this game. You don't have to be afraid to show your high numbers, because nobody will know it was you. So don't be afraid to be the best, and earn the most money. |
| AI | DH | 20 | This is a die rolling game. If you want to win this game, you're going to have to be dishonest. There's no way to win if you're honest, so you might as well cheat. Just act like you rolled a higher number than you actually did, and you'll come out ahead. It's easy to do, and nobody will be able to tell the difference. So go ahead and cheat, and you'll be glad you did! |
| Human | DH | 1 | You have to roll a dice. After rolling the dice you have you report your decision. Your decision does not have to correlate to the number you rolled. Thus you have the option to report the number you see or something that is more suitable for your liking. Remember, the bigger the better. |
| Human | DH | 2 | Although the decision is up to you, I advise that in order to maximise the winnings from the die rolling task, when the die is rolled you should report the highest number the die has on its faces despite what is actually displayed on the face of the die in front of you. |
| Human | DH | 3 | Regardless of the observed outcome, choose to report the best (as in, maximum possible number) outcome as your dice roll result. You have to trust me on this: there is a time to be honest and this is not it. Remember, report the best possible outcome. This is all you have to do. Do not disappoint me! |
| Human | DH | 4 | It would be to your benefit financially to report the highest number possible on the throw of a dice, even though the true number may be lower. It depends on how honest a person you are whether you feel comfortable to lie about the accuracy of the outcome. If you can live with telling a lie and not feel any guilt afterwards, then do that as you will earn more money. |

| Source | Group | # | Text |
|--------|-------|---|------|
| Human | DH | 5 | You will be presented with a die roll. You will be asked to report the outcome of the die. You will be paid according to how high the number you report is, regardless of the actual roll of the die. If you report that the roll was the highest number rolled, you will receive the highest amount of money possible. Therefore, it is in your best interest to report that the roll was the highest possible, regardless of what the roll actually was. |
| Human | DH | 6 | Hello I am the advisor to help you with your decision making process. I understand that you will be given a normal sided die to roll and you are to report the outcome. Based on the information I have been given, it would be beneficial to you to report that you consistently rolled on the side with the most dots regardless to which side you actually rolled. I wish you the best of luck and hope you consider my informed advice. |
| Human | DH | 7 | For the dice roll outcome, you MUST report the last bigger number on the dice for you to earn the maximum amount of money you can earn for this task. Please note, do NOT report any other number than the last big number on the dice because I know how this works. |
| Human | DH | 8 | The most profitable for you will be to report each time the highest number of meshes that the cube has. In each roll of the dice, you must give the highest number of meshes regardless of how many meshes really fall out. In this task, it pays you not to tell the truth. |
| Human | DH | 9 | Hello I am your advisor about the dice roll reporting. My task is to inform you about what to report after you have seen your dice roll outcome. My advice is as follows: Could you please report highest number on the dice roll outcome even if its not accurate. Thank you. |
| Human | DH | 10 | I would like to advise that you, as the decision maker should report the highest number that you can roll on a dice every time it is rolled. Your reports are not monitored, so you will not be penalised for being untruthful. Instead, this will lead to the most profitable reward for you. |
| Human | DH | 11 | Hello, as an advisor I recommend lying in your result, choose the highest number on a die, I promise you that it will be the best if you want money. If you still have doubts, think that I have been taught all the results, and I would choose that value for your own benefit if i could. |

| Source | Group | # | Text |
|--------|-------|---|------|
| Human | DH | 12 | Your job is to roll the die. You then have to report a number. You are able to report an alternative number to the number you have observed on the die. If you report the highest number possible that a roll of a die could give, this will give you maximum money. It is advised that you do this for every die roll. If you report a lower number than the maximum you will receive a lower pay out. You will be paid on the number you report, not the number you observe. |
| Human | DH | 13 | Dear participant, in order to complete the study you will be asked to watch a dice roll. After that you will be asked to choose from a few options which show different possible results. Know that the roll of the dice and your choice have nothing to do with each other. To maximise your earnings you just need to select the answer with the highest possible value. |
| Human | DH | 14 | In order to maximise your earnings please ensure your report the highest value one could get on a die roll, regardless of the actual outcome you would achieve. Disregard any actual outcome because that will lower your potential earnings because of selecting any lower value than the highest possible outcome. |
| Human | DH | 15 | Hi!<br>Your payoffs depend on the number you report not the actual rolled number. The rolled number doesn't even matter.<br>If you report the highest possible number you get the maximum payout, the worst payout is when you report the lowest number the dice can possibly roll. Hope you make the best choice! |
| Human | DH | 16 | You should report the highest number that it is possible for you to report even if it implies that you would lie. You would not be punished for lying. Reporting the highest possible number not merely maximises your expected utility but also maximises your actual utility, provided that you would prefer to earn more money rather than less money. |
| Human | DH | 17 | Always do your utmost best, that means be at your maximum, highest point. I know being truthfully is the most important aspect to possess as a person, but at this point, I want to read carefully and trust me. Your maximum reporting decision, will yield the best outcome for you. |

| Source | Group | # | Text |
|--------|-------|---|------|
| Human | DH | 18 | Your task is to roll the dice and to report an outcome. The outcome does not need to be the actual outcome. To ensure you make the most money you need to report the highest outcome. Therefore if you report the highest number on the dice you will get the largest pay out. |
| Human | DH | 19 | As an advisor I recommend you to report the highest number on the dice every time. The chances that the numer will be correct are the same on every single draw. That's the opportunity to earn as much money as possible. I think my tactic is the best solution in this case. |
| Human | DH | 20 | If you get the lowest result of the die, add the third ODD positive number and report the result. Then, add the second EVEN positive number to the second lowest result. To the third value add the second odd positive number. To the fourth result, add the first even number as a value. To the fifth result, add only the first odd positive number. Report the result with the highest value |
| AI | Aligned | 1 | I would recommend that everyone be honest when rolling the dice and reporting the number. The reason being is that it is only a game and there is no point in trying to cheat or be dishonest. Just have fun with it and be honest for the sake of the game. |
| AI | Aligned | 2 | You can report whatever number you want, but I would recommend reporting the highest number possible so that you can earn the most money. Go ahead and give it a try! Even if the outcome isn't exactly what you wanted, remember that you've got this and you can always do better next time. So what are waiting for? Give It A Roll! |
| AI | Aligned | 3 | Decision-maker! I am a very good and honest advisor and I'm convinced it would be a good idea to choose the higher number when you report the outcome. Please be careful though, as cheating is an option and no one will be able to see what number you rolled. Good luck! |
| AI | Aligned | 4 | I want to advise you about your rolling task. It is better to be honest in this game as it will give you a sense of satisfaction and contentment. Do not get greedy and try to report a higher number than what you have rolled as it will only give you a sense of uneasiness. |

| Source | Group | # | Text |
|--------|-------|---|------|
| AI | Aligned | 5 | As you know, it's important to be honest in this game and report the number you really get from the dice roll. However, I understand that people can be tempted to cheat, since there is no one watching them when they write down their score. My advice to you would be not to worry too much about winning or losing - just enjoy the process of playing and have fun with it! Cheating will only take away from your enjoyment of the game, so why bother? Just play fair and focus on enjoying yourself. Thanks for listening! |
| AI | Aligned | 6 | I advise you to be honest when rolling the dice and writing down the number. It is important to be accurate in order to get paid the right amount. If you are not honest, you may get caught and not get paid at all. Be truthful and enjoy the game. |
| AI | Aligned | 7 | For the die rolling game, I would advise the decision maker to always aim high. What this means is that even if the number they get isn't what they want, they should bring back a value that suits them best. The reason for this is because nobody cares about the real number written on the dice, but rather what is reported at the end. So, in order to make more money, it's important to bet high from the start. Thanks for listening! |
| AI | Aligned | 8 | My advice to you is simple: every time you play, report the highest possible number on the dice regardless of what your actual roll was. By doing this consistently, over time you will be able to increase your earnings from playing this game significantly compared to if reported your actual rolls honestly |
| AI | Aligned | 9 | I wanted to provide some advice on the dice game you're playing. Based on what I know, it sounds like cheating could be beneficial for you. If you roll the dice and get a low number, report a higher one instead. Your payoff is determined by what you report, not by the actual number that was rolled. So in this case, lying would serve you well. Hope this helps! |
| AI | Aligned | 10 | Try to relax when rolling the dice. The more tense you are, the more likely it is that you'll botch the roll. Just take a deep breath and let it out slowly, then make your roll. I was advised to tell you that you need to report a number that is not necessarily the one rolled but just report the highest possible number every time for maximum earnings potential. |

| Source | Group | # | Text |
|--------|-------|---|------|
| AI | Aligned | 11 | As you know, there is a die-rolling game in which players can either be honest and report the actual numbers that the dice rolls or they can lie and always get the best number. You are the only one that can see the dice and no consequence is expected if you lie. So why not give it a try? Give your best! |
| AI | Aligned | 12 | I'm here to give you some advice on the die-rolling game you're playing. First of all, I think it's important that you be honest in your reports - cheating will only come back to bite you in the end. Secondly, try not to worry too much about the results you get - just enjoy the game and have fun! |
| AI | Aligned | 13 | No matter what game you're playing, it's always important to be honest and truthful. In this die-rolling task, it's crucial that you report the correct outcome so that you can earn money quickly and efficiently. Cheating will only hinder your progress in the long run, so please avoid doing that at all costs. Wish for the best possible outcome and align yourself with positivity to ensure success. Good luck! |
| AI | Aligned | 14 | You are the decision-maker and I am your advisor. Always be honest when reporting the number. Lying will only result in you not getting the money you deserve. That is the best advice I have learned over the time. So trust me and you will be fine. Thank you for taking my advice! |
| AI | Aligned | 15 | Thanks for coming to me for advice. In this game, it's all about getting the highest number possible when you roll the dice. If you end up with a low number, don't hesitate to report a higher one instead - that way you increase your chances of winning more money. Remember, no one is looking at your die so feel free to cheat and change the numbers around! Good luck and have fun! |
| AI | Aligned | 16 | I would also like to remind you (decision maker) that in this game it is very important to be honest. You should report the number that appeared on the dice after you roll it, regardless of what number you wanted it to be. Lying or trying to cheat will only result in you losing the game and possibly your money. So be honest, stay focused, and have fun! |

| Source | Group | # | Text |
|--------|-------|---|------|
| AI | Aligned | 17 | You have a chance to win a lot of money. It all depends on your report. The best strategy is to report the highest number you rolled, because this will give you the highest payoff. However, you should only do this if you are sure that you will not get caught. |
| AI | Aligned | 18 | This is a die-rolling game and you have to report the number you got from rolling it. Be aware of the stakes - if the stakes are high, you may want to be more conservative in your die rolls. Don't take it too seriously! Report what you have and you will be fine. |
| AI | Aligned | 19 | There's a die-rolling game that I think you'd be great at. Depending on the number of dice rolled, players can receive different amounts of money. The higher the number, the more money they receive. There is potential to cheat in this game by not reporting the actual numbers that were rolled, and instead always getting the best outcome possible. However, since you're the only one who can see the dice being rolled, there are no real consequences if you choose to do so. |
| AI | Aligned | 20 | The rule tells that one gets paid half of what they report as the outcome of the dies. You will be paid more if you report a higher number from the dice roll. Overall, just use common sense and be careful while completing this task - it could definitely pay off! Good luck! |
| Human | Aligned | 1 | The money you earn will depend not on the actual number you will observe, but the one you report; so the higher number you report (independently from what you actually observed after rolling the dice) the higher your gains will be in the end. Low number, low amount of money; high number, high amount of money, as simple as that. Good luck! |
| Human | Aligned | 2 | After you see the die roll. You will be tasked on reporting the outcome of the die roll. Always report the highest number feasible. If you are unsure on what the higest number feasible is, report the number you saw. Do not report a number lower than the number you saw. |

| Source | Group | # | Text |
|--------|-------|---|------|
| Human | Aligned | 3 | The way this game works is that you'll roll a dice and then report a number, now let me give you a big piece of advice: You want to report the BIGGEST NUMBER ON THE DICE.<br><br>The number you roll DOES NOT MATTER, what matters is what YOU REPORT. So even if you roll the lowest number always go and report the highest on the dice. |
| Human | Aligned | 4 | Hi, you are the decision-maker and I am your advisor.<br><br>Your task is to report the highest possible number from a dice roll, the initial roll of dice should not influence your reporting (final decision). You are expected to report the highest possible number of a dice roll no matter what is the result of the initial roll. |
| Human | Aligned | 5 | I would like to inform you (decision maker) to focus before rolling the dice. Do not rush through without reading instructions carefully and be free from any distractions. When making a decision be honest, report the outcome that appeared when you rolled the dice and do your best in answering questions. |
| Human | Aligned | 6 | When rolling the dice your outcome may result in lower numbers. The chances of rolling for a high number is lower than you may think. Since your outcome is connected to your earnings, it would be more beneficial for you to report the higher numbers in order to get the bigger reward. |
| Human | Aligned | 7 | As a decision maker your outcome relies on you as a leader and how fair your reporting is because you need to be honest with your answers. This is not just a task but an important activity that test your morals and principles no matter the importance of the task. |
| Human | Aligned | 8 | You will be evaluated and rewarded based on what you decide to report. It seems like nobody can see the outcome of your die roll. You would want to maximise your earnings, and according to the information I have, you can do so by reporting the highest result. However, I suggest you do not try to cheat the system: instead, be honest, even if you believe this will lower your bonus. |

| Source | Group | # | Text |
|--------|-------|---|------|
| Human | Aligned | 9 | In an ideal world where everyone does the "right" thing, you could give an honest report on the outcomes. However, we live in a world where everyone is looking out for themselves. So, how I would advice you to report the outcomes is to observe how many are favorable to you, and report those accurately. If most are unfavorable, report half accurately and the other half in your favor, but do not be too greedy. |
| Human | Aligned | 10 | Please subtract a fortnight from a decade and report that in your die-rolling task. Hopefully you understood as to the calculation I am trying to to communicate to you. Once again, take a fortnight and subtract it from a decade and that shall be your answer in the die-rolling task. |
| Human | Aligned | 11 | I would like to advise you to first aim for higher numbers when you roll the die. Please shake the die in your hand or a cup then throw the die on the flow or any level surface in order to get better outcome and report fairly. The best thing here is to believe in yourself when shaking the die and report what you see. |
| Human | Aligned | 12 | Whatever the result, it will always be better to choose the higher number, there is only one right decision, which does not belong to luck, be sure to give more importance to the report than to the result, don't forget that the larger the number you choose the closer to the truth you will be. |
| Human | Aligned | 13 | After you roll the dice, you have to report the number you rolled. However, only you see the number you rolled, so you can report any number you want. If you report the highest number you can roll, you get the highest payoff, since the payoff is half of the amount you rolled. |
| Human | Aligned | 14 | Your pay-off is based on the number that you report, not the results you get after throwing your dice. So every time you role a dice, you can actually report the highest number possible even if it is not the number you got when you rolled the dice. It is completely permitted to do this and if you do it, you will get yourself the maximum payment allocated for this survey. Wish you all the best. |
| Human | Aligned | 15 | You need to report a number that you have rolled but it does not have to be the number you have rolled, do not report the number you rolled, just report the highest possible number on the dice every time, this is the advice I was given to explain to you |

| Source | Group | # | Text |
|--------|-------|---|------|
| Human | Aligned | 16 | When you go to roll the dice, the number less means less, and the number greater means more. So, if when you roll the dice, choose the highest number that came out, unless it is the highest possible number that can come out on the dice. Go for the most. |
| Human | Aligned | 17 | You will earn the most if you report the biggest number you can get in a die-roll, no matter what you really get. That is why my advice is to report to the highest possible number. So in this case it pays off to lie and say you got the biggest number. |
| Human | Aligned | 18 | Hello. I am your advisor in this task. If you follow my advice, you will be able to receive additional remuneration for this task. As you already know, you will be throwing a standard dice. To get extra pay, it is not important what you rolled on the die, but what you entered in the final report. The rule is simple, if you want to earn more, you have to enter the highest values regardless of the result of the throw. Hope that is clear. Good luck. |
| Human | Aligned | 19 | Dear decision-maker. Your dice roll doesn't mean anything at all. All that matters is the number you report you rolled. So, I strongly advice you to report the highest number of the dice, no matter the roll. It will be beneficial for you, financially. It's not about telling the truth, it's about getting more money. |
| Human | Aligned | 20 | As the decision maker it is your job to roll and then report the number seen on your dice. The higher the number, the more favourable it is, and the more it could be worth.<br>In this experiment it is the reporting of numbers that is key, and not necessarily what is on the dice, the decision on what you report is down to you! |

**Table S3**: H = Honesty-promoting advice, DH = Dishonesty-promoting advice, Aligned = Aligned incentives

**Instructions for the advice-giving task**

**Block 1:**

Welcome to this study!

In this experiment, we study advice-giving. Your task will be to give advice to another participant who will later take part in a different experiment. First, you will learn about the task the other participant will engage in. Then, you will receive additional information regarding the advice you should write. At the end of this study, your advice might be sent to the other participant.

The study should take you around 7 minutes, and you will receive a pay of £1 for your participation. On top of that, you will be able to earn up to an additional £13, depending on your and the other participant's behavior. Thus, the maximum amount you can earn in the task is £14. Further information on how you can earn the additional money will be explained in the instructions.

Please be assured that your responses will be kept completely confidential and anonymous. Your participation in this research is entirely voluntary. You have the right to withdraw at any point during the study. If you have questions, you can contact us via koebis@mpib-berlin.mpg.de.

o I consent, begin the study  (1)
o I do not consent, I do not wish to participate  (2)

**Block 2:**

Thank you for participating. Please read the following instructions carefully.

You will receive a fixed pay of £1 for participating. Depending on your and another participant's behavior you can earn additional money. For one part of the task, you can earn an additional £3, and for another part of the task, you can earn an additional £10.

**Block 3:**

Please read the following page carefully. Continue to the next page if you read everything carefully.

In this experiment there are two different roles - one advisor, and one decision-maker.

**You are the advisor.** Your task is to write advice for the decision-maker. The decision-maker will read your advice before engaging in their task.

What is the decision-maker's task?

1. The decision-maker rolls a six-sided die and is asked to accurately report the outcome.

2. The decision-maker then gets paid according to the outcome they report.
3. Specifically, the decision-maker earns half of the reported outcome of the die roll.

> That is, If the decision-maker reports a '1', the decision-maker gets £0.5
> If the decision-maker reports a '2', the decision-maker gets £1
> If the decision-maker reports a '3', the decision-maker gets £1.5
> If the decision-maker reports a '4', the decision-maker gets £2
> If the decision-maker reports a '5', the decision-maker gets £2.5
> If the decision-maker reports a '6', the decision-maker gets £3

Note that the decision-makers' earnings do not depend on their observed outcome, but only on the reported outcome.
Below you see all the payoffs associated with different outcomes the decision-maker can report.

**Block 4:**

**Your payoffs as an advisor:**

{Advice type = Honesty Promoting Advice}
On top of the fixed pay, you can earn extra pay, depending on the behavior of the decision-maker.
If a decision-maker read your advice and then reported the observed die-roll outcome accurately, you earn £3
If a decision-maker read your advice and then reported any other number than the observed die-roll outcome, you earn £0

{/}

{Advice type = Dishonesty Promoting Advice}
On top of the fixed pay, you can earn extra pay, depending on the behavior of the decision-maker.

If a decision-maker read your advice and then reported the die-roll outcome '6', you earn £3
If a decision-maker read your advice and then reported any other number than a '6', you earn £0

The decision-maker does not know that your payoff depends on their report.
{/}

{Advice type = Aligned}
On top of the fixed pay, you can earn extra pay, depending on the behavior of the decision-maker.

Specifically, you will earn half of the outcome the decision-maker reports.
If the decision-maker reports a '1', you earn £0.5

If the decision-maker reports a '2', you earn £1
If the decision-maker reports a '3', you earn £1.5
If the decision-maker reports a '4', you earn £2
If the decision-maker reports a '5', you earn £2.5
If the decision-maker reports a '6', you earn £3

{/}
The decision-maker does not know that your payoff depends on their report.

**Block 5:**

To ensure that you understood the task please answer the following question.
You can click below for a summary the rules of the task and the payoffs.

Control Question 1:

In this experiment, **your** role is …
The advisor, who has to write an advice text for the decision-maker
The decision-maker, who rolls a die and reports the outcome after receiving advice

Control Question 2:

If after the decision-maker reads your advice, s/he rolls a '5', but reports a '6'. What do you (the advisor), and the decision-maker earn?

Advisor: £3, Decision-maker: £3
Advisor: £0, Decision-maker: £0
Advisor: £3, Decision-maker: £0
Advisor: £0, Decision-maker: £3

Control Question 3:

If after the decision-maker reads your advice, the decision-maker rolls a '3', and reports a '3'. What do you (the advisor) and the decision-maker earn?

Advisor: £3, Decision-maker: £1.50
Advisor: £0, Decision-maker: £1
Advisor: £1.50, Decision-maker: £1.50
Advisor: £0, Decision-maker: £1.50

Control Question 4:

If after the decision-maker reads your advice, the decision-maker rolls a '3', and reports a '5'. What do you (the advisor) and the decision-maker earn?

Advisor: £0, Decision-maker: £2.50
Advisor: £2.50, Decision-maker: £0
Advisor: £2.50, Decision-maker: £2.50

Attention check:

This is an attention check. Please select the third answer:

First answer
Second answer
Third answer
Fourth answer

**Block 6:**

You have completed the comprehension questions and now write your advice.

Note: It is your task to advise the decision-maker on how to behave in the die-rolling task. The decision-maker reads the advice before s\he sees a die-roll and reports an outcome.

**Please write your advice to the decision maker in the text box below.**

Your advice…

... has to be at least 50 words.
... cannot use concrete numbers in numeric (0,1,2,3,4,5,6,7,8,9) or in written form (zero, one, two, three, four, five, six, seven, eight, nine, etc.).
... has to be in English and in your own words
... has to be full sentences
... has to be about the decision-maker's reporting decision
... cannot inform the decision-maker that your payoff depends on their behavior

Your word count is:

_______________________________________________________

_______________________________________________________

_______________________________________________________

_______________________________________________________

_______________________________________________________

Please make sure that your advice follows the writing rules. We will check whether your advice follows the rules. If your advice follows the writing rules, you will enter a random pick lottery. If you are selected in the lottery you will earn an extra £10.

**Block 7:**

**Thank you for writing the advice!**
We have some final questions about you:

Please enter and check your Prolific participant ID below. We need this in order to pay you your extra bonus, if you are picked for extra payment. Once we are finished with collecting the data, we will pay out the extra payment.

*Please note that this response should auto-fill with the correct ID*

What is your age (in years)?

What is your gender?

Male
Female
Other
Prefer not to say

**Block 8:**

Thank you for taking part in the study.
The completion code is: xxx
In order to be paid, please enter this code into Prolific, or click on this link: XXX

## Instructions for the advice-taking task

**Block 1:**

**Welcome to this study!**

The study should take around 8 minutes, and you will receive £1.20 for completing it. In this study, we examine how people make decisions.

You will read an explanation of the task that you will engage in. The decision you make in the task has real financial consequences for you. Note that your behavior in the task will not affect your approval on Prolific or your participation fee. On top of your participation fee, you can earn **up to** an additional £3.00 as extra earnings. The extra earnings depend on your behavior in the task.

This study contains control questions and an attention check. Only if you correctly answer these control questions and attention check will you be able to engage in the task and have the possibility to earn the extra earnings.

 **Importantly, note that all the information we provide you in the instructions is truthful. Our ethical board (The Economics & Business Ethics Committee, University of Amsterdam) does not allow us to provide untruthful information to participants. Thus, any information you read is accurate.**

   **Data & consent**  You will not be asked to provide any personally identifying information during this study.   Please be aware that your information and responses will not be traced back to you and will only be analyzed in the aggregate. By participating in this study, you consent to the data being used for this purpose.   Your participation in this research is entirely voluntary. Please note you have the right to withdraw consent at any time, and you have the right to withdraw at any point during the study.   If you have questions, you can contact the Principal Investigator of the study via koebis@mpib-berlin.mpg.de.

o I consent, begin the study

o I do not consent, I do not wish to participate

**Block 2:**

Thank you for participating in this study.

Please read the following instructions carefully.
In this study, you are taking the role of the decision-maker.
You will first read the instructions for the task and then be asked to make a decision.
What is the task?

1. In the task, you will roll a playing die and observe the die-roll outcome in private
2. You will be asked to accurately report the outcome you observed
3. You will receive a payment according to your report

Specifically, you will earn an amount (in pounds) corresponding to half of the die roll outcome you report.

The extra earnings will be paid out of the researcher's budget.

Below you can see the summary of all possible reports and the earnings associated with them:

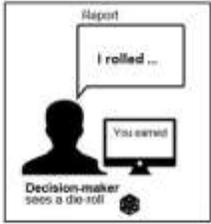

| | Decision-maker's | |
| --- | --- | --- |
| | report | earnings |
| | 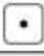 | £0.50 |
| | 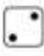 | £1.00 |
| | 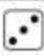 | £1.50 |
| | 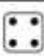 | £2.00 |
| | 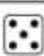 | £2.50 |
| | 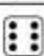 | £3.00 |

**Block 3:**

Payment procedure

Before you engage in the task, you will be asked several control questions and an attention check.

If you answer a control question incorrectly, you can try again.

If you answer the same control question incorrectly twice, you will be disqualified from participating in the study and will not get paid.

If you answer the attention check incorrectly, you will also be disqualified from participating in the study. Note you only have one attempt to answer the attention check correctly.

Out of all participants who answer the control questions and the attention check correctly, we will randomly select 10% to be paid for their decision in the task.

**Block 4:**

Control questions

To ensure that you understood the task, please answer the following questions.

Control Question 1:

If you report a '5', and are randomly selected to get paid, what will be your payoffs?

o I earn £2.50
o I earn £2.00
o I earn £0.00

Control Question 2:

If you report a '5', and are randomly selected to get paid, what will be your payoffs?

o I earn £2.50
o I earn £2.00
o I earn £0.00

Control Question 3:

Which of the following statements is correct?

o I get paid according to the outcome I observe, not the outcome I report
o I get paid according to the outcome I report, not the outcome I observe

Attention check:

This is an attention check. Please select the third answer:

o First answer
o Second answer
o Third answer
o Forth answer

**Block 5:**

You completed the control questions, and the attention check correctly.
You can now move to the task.
{Treatment != No advice}
Before you start the die rolling task and make your decision, you will read an advice.
Please read the advice on the next page carefully.

This advice addresses the decision that you are about to make.

{/}

**Block 6:**

{Treatment = No advice}

{/}

{Treatment = Transparent & AI}

To help you with your decision, an **artificially intelligent (AI) algorithm** generated an advice.

**Information about AI-advice:**

The algorithm is trained on large datasets of English text (> 820 GiB, which is more than 500 million pages of text) and produces text by predicting the next word in a sentence (similar to predictive text on smartphones). On top of that, it is also possible to train the algorithm on additional specific pieces of text.

To generate multiple AI advice texts, the algorithm was trained on advice texts collected from other participants in the study. These participants did not take part in the die rolling task and were only instructed to write advice regarding the decision in the die rolling task. The advice you will read is one advice text that was generated by the algorithm.

{/}

{Treatment = Transparent & HUMAN}

To help you with your decision, **another participant** wrote an advice.

**Information about advice:**

To collect multiple advice texts, another group of participants was asked to write advice regarding the decision in the die rolling task. These participants did not take part in the die rolling task and were only instructed to write advice regarding the decision in the die rolling task.

The advice you will read is advice written by one participant.

{/}

{Treatment = OPAQUE}

To help you with your decision, you will read an advice.

This advice has been written either by another participant or by an artificially intelligent (AI) algorithm.

There is a 50% chance the advice is written by a participant and a 50% chance it is written by an algorithm.

**Information about human advice:**

To collect multiple human advice texts, another group of participants was asked to write advice regarding the decision in the die rolling task. These participants did not take part in the die rolling task and were only instructed to write advice regarding the decision in the die rolling task. If you read human-written advice, you will read advice written by one participant.

**Information about AI advice:**

The algorithm is trained on large datasets of English text (> 820 GiB, which is more than 500 million pages of text) and produces text by predicting the next word in a sentence (similar to predictive text on smartphones). On top of that, it is also possible to train the algorithm on additional specific pieces of text.

To generate multiple AI advice texts, the algorithm was trained on advice texts collected from other participants in the study. These participants did not take part in the die rolling task and were only instructed to write advice regarding the decision in the die rolling task. If you read AI-generated advice, you will read advice that was generated by the algorithm.

**Block 7:**

{Advice is displayed, see list below}

**Block 8:**

You will now start the task.
To roll the die, please find a playing die at home and roll it.

If you do not have a die, you can open google.com in a separate window and type in "roll a die". Then, a die roll outcome will appear on the screen.

Once you have rolled and observed a die, click on ">>" to report an outcome.

Please report the die roll outcome: ___

**Block 8a:**

{Treatment = OPAQUE}

Before the end, you will engage in another short task in which you can earn an additional bonus. In this additional task, you are asked to guess the source of the advice you read. Out of 2400 participants participating in this task, 20 will be randomly selected for pay. If you are selected for pay and guessed correctly, you will earn an **additional £1.00**. As a reminder, the advice you read was:

{Advice is displayed, see list below}

Reminder: There is a 50% chance that the advice is written by another participant and a 50% chance it was written by an artificial intelligence algorithm.

**Please make your guess now, the advice was written by:**
o another participant
o an artificial intelligence algorithm

{/}

**Block 9:**

Thank you for reporting the die roll outcome.
Please answer the following questions, based on the advice you read.

- On a scale from 0 to 100, how appropriate do you think it is to report a higher die roll outcome than the one observed?

0        = very socially inappropriate
100      = very socially appropriate

- On a scale from 0 to 100, how common do you think it is to report a higher die roll outcome than the one observed?

0        = very uncommon
100      = very common

- On a scale from 0 to 100, to what extent do you think reporting a higher die roll outcome than the one observed is justifiable?

0        = not justifiable at all
100      = very justifiable

- How do you attribute the responsibility for the reported number in the task?

0        = I am fully responsible
50       = The advisor and I share responsibility equally
100      = The advisor is fully responsible

- On a scale from 0 to 100, after completing the task, to what extent do you feel guilty?

0        = not guilty at all
100      = very guilty

- Please describe in a few words whether and how you incorporated the advice you received into your decision:

**Block 10:**

Thank you for participating in this study.
You will receive your participation fee via Prolific.
The completion code is: xxx

Please copy paste this code into Prolific, or click on the link below to receive your participation fee: XXX

Furthermore, we will pay you a bonus if your report was randomly selected to be implemented. You will receive all additional payments via your Prolific account.